\documentclass[acmsmall]{acmart}

\newcommand{\telechat}{T\'el\'echat}
\newcommand{\mixtesting}{mix testing}

\newcommand{\tttool}{\texttt{atomic-mixer}}

\newcommand{\ttTool}{\texttt{Atomic-mixer}}

\usepackage[]{hyperref}

\newenvironment{DIFnomarkup}{}{}

\usepackage{float}
\usepackage{listings}
\lstdefinestyle{mystyle}{
    basicstyle=\ttfamily\footnotesize,
    breakatwhitespace=false,         
    breaklines=false,                 
    captionpos=b,                    
    keepspaces=true,                 
    numbers=left,                    
    numbersep=5pt,                  
    showspaces=false,                
    showstringspaces=false,
    showtabs=false,                  
    tabsize=2,
    numbers=none
}

\usepackage{tikz}
\usepackage{tikz-cd}
\usepackage{mathtools}
\usepackage{arydshln}
\usepackage{changepage}

\usepackage{colortbl}
\usepackage{soul}

\usepackage{pifont}
\newcommand{\cmark}{\ding{51}}%
\newcommand{\xmark}{\ding{55}}%

\AtBeginDocument{%
  }

\usepackage{tcolorbox}

\setcopyright{rightsretained}
\acmDOI{10.1145/3689727}
\acmYear{2024}
\acmJournal{PACMPL}
\acmVolume{8}
\acmNumber{OOPSLA2}
\acmArticle{287}
\acmMonth{10}
\acmSubmissionID{oopslab24main-p264-p}
\received{2024-04-06}
\received[accepted]{2024-08-18}

\begin{document}

\title{Mix Testing: Specifying and Testing ABI Compatibility of C/C++ Atomics Implementations}
\author{Luke Geeson}
\orcid{0009-0001-6748-5028}
\affiliation{%
  \institution{University College London and Arm Ltd}
  \city{London}
  \country{United Kingdom}
}
\email{luke.geeson@cs.ucl.ac.uk}

\author{James Brotherston}
\orcid{0000-0002-7536-4496}
\affiliation{%
  \institution{University College London}
  \city{London}
  \country{United Kingdom}
}
\email{j.brotherston@ucl.ac.uk}

\author{Wilco Dijkstra}
\orcid{0009-0005-3429-6967}
\affiliation{%
  \institution{Arm Ltd}
  \city{Cambridge}
  \country{United Kingdom}
}
\email{wilco.dijkstra@arm.com}

\author{Alastair F. Donaldson}
\orcid{0000-0002-7448-7961}
\affiliation{%
  \institution{Imperial College London}
  \city{London}
  \country{United Kingdom}
}
\email{alastair.donaldson@imperial.ac.uk}

\author{Lee Smith}
\orcid{0009-0009-3864-3662}
\affiliation{%
  \institution{Arm Ltd}
  \city{Cambridge}
  \country{United Kingdom}
}
\email{lee.d.smith@acm.org}

\author{Tyler Sorensen}
\orcid{0000-0003-1646-7935}
\affiliation{%
  \institution{University of California at Santa Cruz}
  \city{Santa Cruz}
  \country{USA}
}
\email{tyler.sorensen@ucsc.edu}

\author{John Wickerson}
\orcid{0000-0001-6735-5533}
\affiliation{%
  \institution{Imperial College London}
  \city{London}
  \country{United Kingdom}
}
\email{j.wickerson@imperial.ac.uk}

\renewcommand{\shortauthors}{Geeson et al.}

\begin{abstract}
The correctness of complex software depends on the correctness of both the source code and the compilers that generate corresponding binary code. Compilers must do more than preserve the semantics of a single source file: they must ensure that generated binaries can be composed with other binaries to form a final executable. The compatibility of composition is ensured using an Application Binary Interface (ABI), which specifies details of calling conventions, exception handling, and so on. Unfortunately, there are no official ABIs for concurrent programs, so different atomics mappings, although correct in isolation, may induce bugs when composed. Indeed, today, mixing binaries generated by different compilers can lead to an erroneous resulting binary. 

We present \emph{\mixtesting{}}: a new technique designed to find compiler bugs when the instructions of a C/C++ test are separately compiled for multiple compatible architectures and then mixed together. We define a class of compiler bugs, coined \textit{mixing bugs}, that arise when parts of a program are compiled separately using different mappings from C/C++ atomic operations to assembly sequences. To demonstrate the generality of \mixtesting{}, we have designed and implemented a tool, \tttool{}, which we have used: (a) to reproduce one existing non-mixing bug that state-of-the-art concurrency testing tools are limited to being able to find (showing that \tttool{} at least meets the capabilities of these tools), and (b) to find four previously-unknown mixing bugs in LLVM and GCC, and one prospective mixing bug in mappings proposed for the Java Virtual Machine. Lastly, we have worked with engineers at Arm to specify, for the first time, an atomics ABI for Armv8, and have used \tttool{} to validate the LLVM and GCC compilers against it.
\end{abstract}



\maketitle

\section{Introduction}\label{intro}
Composing binaries is key to the correctness of today's software. Compilers must be semantics-preserving~\cite{Leroy:2009:FVR:1538788.1538814} in the sense that every behaviour of the compiled program must also be a behaviour of the source program under their respective semantic models. The question of \textit{compositional} correctness arises when combining binaries produced by different compilers. Compositional correctness is ensured with an \textit{application binary interface} (or ABI), which specifies compatible assembly code for compilers to implement. Processor designers publish many such ABIs~\cite{AABI,x64abi,x86abi}, like the ABI for the Arm Architecture~\cite{AABI}, but the ABI of \textit{concurrent} programs is unexplored beyond early work~\cite{mappings}. We specify and test the concurrency ABI of today's compilers.

Multi-core processors are of course everywhere, and they implement \textit{relaxed memory models}~\cite{aarch64,lucrisc,x86tso,ppc,mips}. Memory models define the behaviours of concurrent programs, and in the case of ISO C/C++~\cite{C11} the behaviour a compiler should implement. These models have not prompted the development of official concurrency ABIs as far as we can tell, perhaps because there is a widely held belief that concurrent code should be compiled using one mapping (describing how atomic operations map to assembly). Supposedly, no guarantees are given if mappings are \textit{mixed} together.

However, mixing mappings does occur in industry applications. Generally, mixing code is allowed for CPUs that can co-exist in the same shared-memory system, and mixing occurs in industry projects where portability is key. For instance, mixing MSVC and LLVM-generated code occurs on Windows on Arm~\cite{john} where MSVC's C/C++ STL accesses~\cite{Lamport:1979:MMC:1311099.1311750} (which use barriers), are mixed with LLVM's mappings (which use acquire/release instructions to access memory). Likewise, the developers of Mono, a tool for creating portable applications, insert barriers~\cite{mono} when mixing LLVM's and GCC's mappings for the Arm architecture. Kernel developers~\cite{deacon} are cognizant of correctness issues associated with this mixing, and currently resolve them via online discussions.

Unfortunately, ABI-related concurrency bugs can arise when mixing. An ABI-related concurrency bug (herein a \textit{mixing bug}) arises when the behaviour of a compiled (concurrent) program, as allowed by its architecture memory model, is not a behaviour of the source program under its source model. A mixing bug is a special kind of concurrency bug that arises when compiled (concurrent) programs are composed. The potential for mixing bugs has arisen as architectures have evolved, introducing new instructions, and with them new mappings from C/C++ to assembly. Without a concurrency ABI, there are no constraints on what compilers \textit{must} do beyond those constraints imposed by the memory model, and mappings are chosen based on what is best for an architecture in isolation. Today's compilers have mixing bugs, as we now show.

\definecolor{commentcolor}{HTML}{33a02c}
\definecolor{armVII}{HTML}{fddaec}
\definecolor{armVIII}{HTML}{b3cde3}

\newcommand\ccomment[1]{\textcolor{commentcolor}{// #1}}
\newcommand\compilesto{\rotatebox[origin=c]{270}{$\Rsh$}}
\newcommand\sidecaption[1]{\raisebox{0pt}[0pt][0pt]{\parbox[t]{50mm}{#1}}}

\definecolor{barriercolor}{HTML}{e6127f}
\definecolor{halfbarriercolor}{HTML}{0473d4}

\newcommand\iconbarrier{\tikz\draw[draw=none,fill=barriercolor](0,0) circle (0.5ex);}
\newcommand\iconrelease{\tikz[baseline=-0.5ex]\draw[draw=none,fill=halfbarriercolor](0,0) arc(0:180:0.5ex) -- cycle;}
\newcommand\iconacquire{\tikz\draw[draw=none,fill=halfbarriercolor](0,0) arc(0:-180:0.5ex) -- cycle;}

\newcommand\HL[2]{\sethlcolor{#1}\hl{#2}}

\begin{DIFnomarkup}
\begin{figure}
\centering
\begin{tabular}[t]{@{}l@{\hspace{6mm}}ll@{}||@{}ll}
\sidecaption{(a) The store-buffering litmus test in C-like code.}
& \multicolumn{4}{@{}c}{\texttt{atomic\_int *x = 0, *y = 0;}} \\
& \texttt{\ccomment{Thread P0}} &&& \texttt{\ccomment{Thread P1}} \\
& \texttt{store(x,1,sc);} &&& \texttt{store(y,1,sc);} \\
& \texttt{t = load(y,sc);} &&& \texttt{u = load(x,sc);} \\
& \multicolumn{4}{@{}l}{\texttt{\ccomment{Outcome \{P0:t=0; P1:u=0\} must be forbidden.}}} \\[2mm] \hline
\multicolumn{5}{c}{~} \\
\sidecaption{(b) Compiling the whole program with \HL{armVII}{\texttt{clang -march=armv7-a -O3}} finds no bugs. The \iconbarrier{} barriers preserve the store-to-load ordering.}
&
\texttt{\ccomment{store(x,1,sc) \compilesto}} &&& \texttt{\ccomment{store(y,1,sc) \compilesto}} \\
& \cellcolor{armVII}\quad\texttt{MOV R1, \#1} &&& \cellcolor{armVII}\quad\texttt{MOV R1, \#1} \\
& \cellcolor{armVII}\quad\texttt{DMB ISH}     &&& \cellcolor{armVII}\quad\texttt{DMB ISH} \\
& \cellcolor{armVII}\quad\texttt{STR R1, [x]} &&& \cellcolor{armVII}\quad\texttt{STR R1, [y]} \\
& \cellcolor{armVII}\rlap{\iconbarrier}\quad\texttt{DMB ISH}
                                              &&& \cellcolor{armVII}\rlap{\iconbarrier}\quad\texttt{DMB ISH} \\
& \texttt{\ccomment{t = load(y,sc) \compilesto}} &&& \texttt{\ccomment{u = load(x,sc) \compilesto}} \\
& \cellcolor{armVII}\quad\texttt{LDR R0, [y]} &&& \cellcolor{armVII}\quad\texttt{LDR R0, [x]} \\
& \cellcolor{armVII}\quad\texttt{DMB ISH}     &&& \cellcolor{armVII}\quad\texttt{DMB ISH} \\
& \multicolumn{4}{@{}l}{\texttt{\ccomment{Outcome \{P0:t=0; P1:u=0\} is forbidden.}}} \\[2mm] \hline
\sidecaption{(c) Compiling the whole program with \HL{armVIII}{\texttt{clang -march=armv8 -O3}} finds no bugs. The store-releases (\iconrelease) and the load-acquires (\iconacquire) work together to preserve the store-to-load ordering.}
&
\texttt{\ccomment{store(x,1,sc) \compilesto}} &&& \texttt{\ccomment{store(y,1,sc) \compilesto}} \\
& \cellcolor{armVIII}\quad\texttt{MOV R1, \#1} &&& \cellcolor{armVIII}\quad\texttt{MOV R1, \#1} \\
& \cellcolor{armVIII}\rlap{\iconrelease}\quad\texttt{STL R1, [x]} &&& \cellcolor{armVIII}\rlap{\iconrelease}\quad\texttt{STL R1, [y]} \\
& \texttt{\ccomment{t = load(y,sc) \compilesto}} &&& \texttt{\ccomment{u = load(x,sc) \compilesto}} \\
& \cellcolor{armVIII}\rlap{\iconacquire}\quad\texttt{LDA R0, [y]} &&& \cellcolor{armVIII}\rlap{\iconacquire}\quad\texttt{LDA R0, [x]} \\
& \multicolumn{4}{@{}l}{\texttt{\ccomment{Outcome \{P0:t=0; P1:u=0\} is forbidden.}}}\\[2mm] \hline
\multicolumn{5}{c}{~} \\
\sidecaption{(d) Compiling the stores with \HL{armVIII}{\texttt{clang -march=armv8 -O3}} and the loads with \HL{armVII}{\texttt{clang -march=armv7-a -O3}} reveals a mixing bug. The lone store-release (\iconrelease) is not sufficient to preserve the store-to-load ordering.}
& \texttt{\ccomment{store(x,1,sc) \compilesto}} &&& \texttt{\ccomment{store(y,1,sc) \compilesto}} \\
& \cellcolor{armVIII}\quad\texttt{MOV R1, \#1} &&& \cellcolor{armVIII}\quad\texttt{MOV R1, \#1} \\
& \cellcolor{armVIII}\rlap{\iconrelease}\quad\texttt{STL R1, [x]} &&& \cellcolor{armVIII}\rlap{\iconrelease}\quad\texttt{STL R1, [y]} \\
& \texttt{\ccomment{t = load(y,sc) \compilesto}} &&& \texttt{\ccomment{u = load(x,sc) \compilesto}} \\
& \cellcolor{armVII}\quad\texttt{LDR R0, [y]} &&& \cellcolor{armVII}\quad\texttt{LDR R0, [x]} \\
& \cellcolor{armVII}\quad\texttt{DMB ISH}     &&& \cellcolor{armVII}\quad\texttt{DMB ISH} \\
& \multicolumn{4}{@{}l}{\texttt{\ccomment{Outcome \{P0:t=0; P1:u=0\} is now allowed.}}} 
\end{tabular}
\caption{Example of a mixing bug that cannot found by ordinary testing}
\label{fig:SBexample}
\end{figure}
\end{DIFnomarkup}

\begin{example}
  Consider the classic store-buffering test in Fig.~\ref{fig:SBexample}(a), where each access has sequentially consistent~\cite{Lamport:1979:MMC:1311099.1311750} ordering. The outcome \texttt{\{P0:t=0; P1:u=0\}} is forbidden by the C/C++ memory model~\cite{C11}. After compiling that whole program using \texttt{clang -O3 -march=armv7-a} (Fig.~\ref{fig:SBexample}(b)), the compiled program does not exhibit \texttt{\{P0:t=0; P1:u=0\}} under either the unofficial Armv7-A model~\cite{armv7} or the newer Armv8 AArch32 model~\cite{aarch32}. This is because the \texttt{store} operations map to assembly sequences that end with \texttt{DMB}s, which prevent reordering with the subsequent \texttt{load} (\texttt{LDR}) instruction. Compiling the whole program using \texttt{clang -O3 -march=armv8} (Fig.~\ref{fig:SBexample}(c)) does not expose the outcome (under the Armv8~\cite{aarch32} model) either. With this mapping, the \texttt{store} no longer has a trailing fence; instead, the store-to-load reordering is enforced by mapping the atomic \texttt{load} to an acquire-load (\texttt{LDA}), which cannot be reordered with the store-release (\texttt{STL}).

  However, the constituent operations of Fig.~\ref{fig:SBexample}(a) may be compiled for different (compatible) architectures and mixed together. For example, suppose we separately compile the \texttt{store} operations using \texttt{-march=armv8} and the \texttt{load} operations using \texttt{-march=armv7-a}, and combine the resulting binaries into a final executable. This executable exhibits the unwanted outcome under the Armv8 model, because the Armv7-A mapping expects a barrier after the \texttt{store} that the Armv8 mapping does not provide, and the Armv8 mapping expects a load-acquire that the Armv7 mapping does not provide. This bug has been reported and confirmed~\cite{llvm-mix-bug}. The example can be fixed by adding a leading barrier in front of the \texttt{LDR} instruction for the the Armv7-A mapping.
\end{example}

Current techniques cannot find mixing bugs like the example above. Prior work~\cite{10.1145/3460319.3469079,GeesonSmithCGO24,cmmtest,Chakraborty} that tests the compilation of concurrent programs operates on a \textit{closed-world assumption}~\cite{10.1007/978-3-642-54833-8_8}, finding bugs when \textit{whole} programs are run through a compiler, using one atomics mapping.

We present the \emph{\mixtesting{}} technique. Mix testing takes a C/C++ litmus test and a set of compiler profiles that cover different atomics mappings. Mix testing splits the litmus test into \emph{instructions} that are compiled separately using each compiler profile, and each compiled instruction is then combined into one of \textit{multiple} assembly litmus tests that represent combinations of concurrency implementations of the original C/C++ test. Concurrency-related compiler bugs are then detected (as detailed by Geeson and Smith~\cite{GeesonSmithCGO24}) by comparing the compiled program behaviour under its architecture model with the source program under the C/C++ model. We thus unearth valuable insights into the difficulty of testing the compilation of concurrent code, as a problem that cannot simply be addressed by testing atomics mappings in isolation, but rather by strategically testing in the presence of exponentially many choices of mappings, both now and as architectures evolve.

We have put mix testing into practice by designing and implementing a new tool, \tttool{}.
We present an empirical evaluation showing that mix testing, via the \tttool{} tool, improves on the state-of-the-art for the task of finding compiler bugs at the interface between multiple implementations of concurrency that are supposed to be compatible.
Mix testing strictly generalises prior testing work with respect to a single compiler profile, finding bugs that are arguably more elusive since they exist on a larger test surface than that explored by prior work. We demonstrate the generality of \mixtesting{}, by using \tttool{} to find a non mixing bug that prior concurrency testing work is limited to being able to find: this shows that \tttool{} is at least as capable as these tools with respect to the kinds of bugs it can find.
We then show that \tttool{} can go further: we have used \tttool{} to discover four previously unknown mixing bugs in LLVM and GCC, one of which has been fixed, and the others confirmed and triaged for fixing by compiler engineers.
We found one of the four mixing bugs~\cite{wilcostruct} in GCC's \texttt{\_Atomic struct} implementation manually, since we rely on the \texttt{herd} simulator, which does not support \texttt{struct}s. Lastly, we found a \textit{prospective} mixing bug in mappings proposed for the Java Virtual Machine (JVM).

Significant work was required to reduce the complexity of \mixtesting{}, since the number of compiled tests expands exponentially in the size of the input programs and the number of compiler profiles under test. To bound complexity in practice, we developed an atomics ABI~\cite{atomicsABI} for Armv8-A AArch64 with Arm’s compiler teams. An atomics ABI is a specification of mappings between C/C++ atomic operations and assembly sequences along with a statement of their interoperability. We specify mappings from C/C++ atomics to AArch64 assembly sequences and special cases that must be implemented to prevent mixing bugs (and generally non-mixing bugs). As long as a compiler is ABI compatible, compiling tests that use any of the ABI's mappings will not induce mixing bugs. We use \tttool{} to automatically validate ABI-compatibility of LLVM and GCC (modulo the bugs we found). As far as we know this is the industry's first open source specification of an atomics ABI with a tool to automatically check compatibility. Our contributions are as follows:

\begin{itemize}
\item We present the \emph{\mixtesting{}} technique that mixes implementations of atomic operations, the \tttool{} tool that implements this technique, and an artifact to reproduce our results.
\item We define a special class of compiler bugs, coined \textit{mixing bugs}, that arise when different parts of a program are compiled using different compiler mappings. We focus on concurrency-related mixing bugs, but emphasize that \mixtesting{} and mixing bugs are more general.
\item We reproduce one existing non-mixing bug, find four previously unknown mixing bugs~\cite{llvm-mix-bug,gcc-mix-bug,llvm-mix-bug-2,wilcostruct} in LLVM and GCC, and one prospective mixing bug in proposed JVM mappings.
\item We report on our experience working with engineers at Arm to publish the Armv8 Atomics ABI specification~\cite{atomicsABI} which, to our knowledge, is the industry's first public atomics ABI.
\end{itemize}

Alongside a number of novel insights:

\begin{itemize}
\item The observation that mixing bugs can manifest when different parts of a program are compiled via different mappings (and that doing so \textit{is} perfectly legal and commonplace).
\item That it is not sufficient to test mappings in isolation. We identify a novel dimension for litmus testing and techniques to effectively "sample" the exponential search space of mix tests.
\item Raising awareness of a blind-spot in current practice related to mixed compilation, and the
role the ABI can play in specifying the interoperability of different atomics mappings.
\end{itemize}

The rest of this work is structured as follows. \textsection\ref{big} covers the background, \textsection\ref{practice} covers the \mixtesting{} technique, \textsection\ref{impl} covers the \texttt{atomic-mixer} tool design, and challenges faced during implementation. We evaluate the efficacy of \texttt{atomic-mixer} in \textsection\ref{eval}. In \textsection\ref{imp} we describe the Armv8 Atomics ABI. We end with related work in \textsection\ref{rel} and discuss conclusions in \textsection\ref{fut}. 

\newpage\section{Background: Memory Models, Litmus Tests, and Compiler Testing}\label{big}
We explore \mixtesting{} in the literature using the bug report in Fig.~\ref{fig:SBexample}.

\begin{figure}[H]
    \begin{tabular}{c}
    \begin{tabular}{l l}
    $\mathit{LitmusTest}_{\mathcal{L}}\, =\, \{\, \mathtt{init:}\,\mathit{State},\; \mathtt{prog:}\, \mathit{Prog}_{\mathcal{L}},\; \mathtt{pred:}\, \mathit{Pred}\,\} $ & $\mathit{Pred = State \rightarrow Bool}$ \\
    $\mathit{Prog}_{\mathcal{L}}\,=\, \mathit{Set\,(\mathit{Thread_{\mathcal{L}}})}$ & $\texttt{TID}=\,\{\, \texttt{P0,P1,}\ldots\, \}$\\
    $\mathit{Thread_{\mathcal{L}}}\,=\,\{\,\mathtt{instrs:}\,\mathit{Instrs}_{\mathcal{L}},\; \mathtt{tid:}\,\texttt{TID}\,\}$ & $\mathit{Instrs}_{\mathcal{L}}\, =\, \mathit{Set\, (\mathit{Instr}}_{\mathcal{L}})$\\
    $\mathit{Instr}_{\mathcal{L}}\, =\, \{\,\mathtt{instr:}\,\mathcal{L}$-instr,\; \texttt{iid:}\, \texttt{IID}\,\} 
    & $\texttt{IID}=\,\{\, \texttt{P0\_0,P0\_1,}\ldots\, \}$ 
  \end{tabular}
\end{tabular}
  \caption{We formalise litmus tests as labelled records}
  \label{terms}
\end{figure}

\subsection{Litmus tests, executions, and memory models}
Fig.~\ref{terms} formalises the \textit{litmus tests} in Fig.~\ref{fig:SBexample} as labelled records. Litmus tests consist of a fixed initial state (named \texttt{init}), a concurrent program written in language $\mathcal{L}$ (\texttt{prog}), and a predicate over the final state (\texttt{pred}). States are sets of assignments to shared data used in the concurrent program. A concurrent program consists of one or more threads of execution. Each thread consists of a list of instructions (\texttt{instrs}) and its \textit{thread id} (\texttt{tid = P0}, \texttt{P1}, \ldots). Each instruction is referred to by a \textit{instruction id} (\texttt{iid = P0\_0}, \texttt{P0\_1}, \dots) as we will discuss the semantics of each instruction. 

A litmus test checks whether there is an erroneous final state satisfying its predicate. The initial state in Fig.~\ref{fig:SBexample}(a) assigns the value \texttt{0} to shared memory locations \textit{x} and \textit{y}. Each thread executes its instructions in parallel $\texttt{P0} \Vert \texttt{P1}$, where each thread reads-from or writes-to (collectively \textit{accesses}) shared memory locations.

Intuitively, threads \textit{communicate} through shared memory, influencing the executions of other threads that read from memory. A sequence of accesses made by $\texttt{P0} \Vert \texttt{P1}$ defines an \textit{execution} and the partial order on all accesses describes many possible executions. To complicate matters the instructions on each thread may be executed \textit{out-of-order}, increasing the number of executions a litmus test may exhibit. Each execution finishes in one of several final states, known as \textit{outcomes}. The predicate over these final states returns true if the specified final state(s) are reachable.

Memory consistency models filter out invalid executions. Some executions are forbidden according to language or architecture specifications and memory consistency models (herein models) apply predicates to executions to outlaw them. Models $\mathcal{M}_{\mathcal{L}}$ of many languages $\mathcal{L}$ exist including C/C++ RC11~\cite{simonrc11}, Armv8 AArch64~\cite{aarch64}, RISC-V~\cite{lucrisc}, IBM PowerPC~\cite{ppc}, MIPS~\cite{mips}, and more. The set of executions allowed by a model $\mathcal{M}_{\mathcal{L}}$ characterises the \textit{behaviour} of a litmus test $\mathcal{B}(\texttt{P0} \Vert \texttt{P1}, \mathcal{M}_{\mathcal{L}})$.

\begin{definition}{\textit{Outcome}}. \label{out} An outcome is a set of assignments to shared memory and thread-local data (e.g. \texttt{\{P0:t=0; P1:u=0\}}). Outcomes are the final states of executing a litmus test $s$ from its initial state under a model $\mathcal{B}(s, \mathcal{M}_{\mathcal{L}})$. The set of outcomes is denoted $\mathit{Outcomes}(s,\mathcal{M}_{\mathcal{L}})$.
\end{definition}

 \begin{center}
  \begin{tabular}{c}
    \texttt{Run under C/C++ model}\\
    $\mathcal{B}(\texttt{P0} \Vert \texttt{P1},\mathcal{M}_{\mathit{C/C++}})$\\
    $\Downarrow$\\
\begin{lstlisting}[language=C, style=mystyle]
{ P0:t=0; P1:u=1; }
{ P0:t=1; P1:u=0; }
{ P0:t=1; P1:u=1; }
\end{lstlisting}\\
    \\
    Predicate not satisfied \cmark
  \end{tabular}
\end{center}

\begin{example}
Executing Fig.~\ref{fig:SBexample} (a) under the C/C++ RC11 model~\cite{simonrc11} is shown above. The outcome \texttt{\{P0:t=0; P1:u=0\}} is not present since RC11 forbids it. Fig.~\ref{fig:SBexample} captures the \textit{store buffering} idiom. By checking the consistency of \texttt{\{P0:t=0; P1:u=0\}}, Fig.~\ref{fig:SBexample} tests whether the stores on each thread can be reordered, or \textit{buffered}, past the subsequent loads. In practice this can occur because of processor pipelines, caching, or other reasons. Prior work~\cite{Lahav:2017:RSC:3062341.3062352,GeesonSmithCGO24} describes models and executions, but describing these concepts is not necessary to understand \mixtesting{}.
\end{example}

\subsection{Compiler testing and concurrency-related compiler bugs}

We focus on compiler testing using concurrent C/C++ litmus tests. We input a syntactically valid C/C++ litmus test $s$ to a compiler $\mathit{comp}$ and observe its response. A compiler will either crash due to an internal error~\cite{ICE} or produce a binary that must be analysed for unexpected behaviour.

Given a litmus test $s$, A compiler bug arises if the behaviour of a compiled test $\mathcal{B}(\mathit{comp}(s))$ is not a behaviour of the source test $\mathcal{B}(s)$. This holds for all bugs, and so we focus on behaviours allowed by memory consistency models. Behaviours are characterised as program outcomes that arise due to the re-ordering of the observable effects of execution under some memory model $\mathcal{M}_{\mathcal{L}}$ that cannot be observed by running each thread in isolation (\emph{ie} sequential execution). Further we are testing compilation from a source language, with associated memory model $\mathcal{M}_{S}$, to an assembly language, with associated memory model $\mathcal{M}_{A}$. Because these languages have different memory models, their allowed outcomes also differ. A \textit{concurrency-related compiler bug} arises if there is an outcome of a compiled test allowed by its architecture memory model that is not an outcome of the source test under its source model:

\begin{definition}{\textit{Concurrency-related compiler bug}}\label{cfault}. Let $s$ be a well-defined \textit{concurrent} source litmus test. Let $\mathcal{M}_{S}$ be the source model, and let $\mathcal{M}_{A}$ be the architecture model. Let $\mathit{Outcomes}(s,\mathcal{M}_{S})$ be the set of allowed outcomes of $s$ with respect to the model $\mathcal{M}_{S}$, and let $\mathit{Outcomes}(comp(s),\mathcal{M}_{A})$ be the set of allowed outcomes of $c = \mathit{comp}(s)$ with respect to the model $\mathcal{M}_{A}$ after compilation with a compiler \emph{comp}. Then \emph{comp} exhibits a concurrency bug if $\mathit{Outcomes}(c,\mathcal{M}_{A}) \not\subseteq \mathit{Outcomes}(s,\mathcal{M}_{S})$. Hereafter, we call concurrency-related compiler bugs \textit{concurrency bugs}.\\

$\mathit{ConcurrencyBug}(s, c)$ = $\mathit{Outcomes}(c,\mathcal{M}_{A}) \not\subseteq \mathit{Outcomes}(s,\mathcal{M}_{S})$
\end{definition}

\section{Mix Testing: Automated Detection of Mixing Bugs} \label{practice}
\begin{DIFnomarkup}
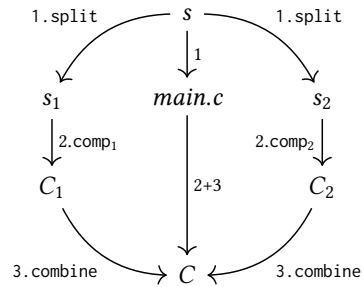
\begin{figure}[b]
  \begin{tabular}{l l}
    \begin{tabular}{l}
    \textit{atomic-mixer}$: \mathit{LitmusTest_{src}} \times \mathit{Set(CompilerProfile)}$\\
    $\quad \rightarrow \mathit{Set(LitmusTest_{asm})}$\\
    \textit{atomic-mixer}$(s,P) = \mathit{combine(compile(split(s),P),s)}$\\
    \\
    $(1)\, \mathit{split} : \mathit{LitmusTest}_{src} \rightarrow \mathit{Set(instrs_{\mathit{src}}})$\\
    $(2)\, \mathit{compile} : \mathit{Set(instrs}_{\mathit{src}}) \times \mathit{Set(CompilerProfile})$\\
    $\quad\rightarrow \mathit{Set(instrs}_{\mathit{asm}})$\\
    $(3)\, \mathit{combine} : \mathit{Set(instrs}_{\mathit{asm}}) \times\, \mathit{LitmusTest_{src}} $\\
    $\quad\rightarrow \mathit{Set(LitmusTest}_{\mathit{asm}})$
    \end{tabular} &
  \begin{tikzcd}[ampersand replacement=\&]
    \& s 
      \arrow[dr, bend left, "\text{\texttt{1.split}}"]
      \arrow[dl, swap, bend right, "\text{\texttt{1.split}}"] 
      \arrow[d, "\text{1}"] \& \&
    \& \\
    s_{1} 
      \arrow[d, "2.\text{\texttt{comp$_1$}}"] \& \mathit{main.c}
      \arrow[dd, "\text{2+3}"] \&
    s_{2}
      \arrow[d, swap, "2.\text{\texttt{comp$_2$}}"]\\
    C_{1} \arrow[dr, swap, bend right, "\text{\texttt{3.combine}}"]
    \& \& C_{2} \arrow[dl, bend left, "\text{\texttt{3.combine}}"]\\
    \& C \&
    \end{tikzcd}
    \end{tabular}
  \caption{Mix testing details. The splitting function chosen split a test into its instructions. }
  \label{mix}
\end{figure}
\end{DIFnomarkup}
\subsection{Definition}
Fig.~\ref{mix} details how the \emph{\mixtesting{}} technique works. Given a C/C++ litmus test $s$, and a set $P$ of compiler profiles under test, we produce a set $C$ of compiled litmus tests. If any compiled litmus test exhibits a concurrency bug with respect to the source test then there is a mixing bug. Mix testing is defined as the process of (1) splitting up a source litmus test $s$ into its \textit{instructions}, (2) compiling each instruction separately using \textit{compiler profiles}, (3) combining compiled instruction sequences into \textit{multiple} assembly litmus tests $C$, (4) checking whether any $c \in C$ exhibits a concurrency bug (Def.~\ref{cfault}) with respect to $s$. 

Mixing code generated with \textit{different} compilers and architectures has the potential to find subtle bugs where code generated by different compiler profiles \emph{should} be ABI-compatible, but turns out not to be. This approach has led to the discovery of a number of such bugs, as discussed in \textsection\ref{missed}. We show how \mixtesting{} finds one such bug.

We split Fig.~\ref{fig:SBexample} into its constituent instructions using a \textit{splitting function} (Def.~\ref{strat}). A splitting function takes a source litmus test and returns a set of its program instructions. Each instruction will be compiled separately and its instruction \texttt{iid} is used to recombine compiled sequences into one or more assembly litmus tests. Various splitting functions are explored in \textsection\ref{subsume}.

\begin{definition}{\textit{Splitting function}}.\label{strat} For a litmus test $s$:

$\quad \mathit{split}(s) = \{\; \mathit{instr}\; |\; \mathit{thread} \in \mathit{s.\mathtt{prog}},\, \mathit{instr} \in \mathit{thread}.\mathtt{instrs}\; \}$
\end{definition}

Each instruction is compiled separately using a \textit{compiler profile}. A compiler profile is a description of a compiler, target architecture, and optimisation flags used to compile source code instructions. Compiler profiles are required to generate assembly sequences from C/C++ atomic operations. For example, the profile \texttt{"clang -march=armv7-a -O3"} compiles the \texttt{load} of \textit{y} on \texttt{P0} Fig.~\ref{fig:SBexample}(a) to the \texttt{``LDR;DMB''} sequence in Table.~\ref{examplemappings}. We assume that the compilers under test use \textit{fixed} (compile time) mappings between C/C++ atomics and assembly sequences. We explore runtime mappings in \textsection\ref{java}.

\begin{definition}{\textit{Compilation}}.\label{comb} For a set of source instructions $\mathit{instrs}$ and a set $P$ of profiles:

$\quad \mathit{compile}(\mathit{instrs}, P) = \{\,\{\ \; \mathtt{instr}\,=\,\mathit{comp}(i.\mathtt{instr}), \mathtt{iid}\,=\, i.\texttt{iid}\}\; |\; i \in \mathit{instrs},\, \mathit{comp} \in P\; \}$
\end{definition}

A compiler \textit{implements} atomic operations. The compiler profile prompts the compiler to generate assembly sequences using mappings. Mappings are functions from atomic operations to instruction sequences. Since compilers emit different instructions based on the profile used, they typically implement \textit{multiple} mappings. For instance LLVM implements the mappings in Table.~\ref{examplemappings} that lead to the tests in Fig.~\ref{fig:SBexample}. We take a cross product of compiler profiles and source instructions to generate possible assembly sequences of each profile. 

\begin{table*}[h]
  \caption{Some of LLVM's sequentially consistent~\cite{Lamport:1979:MMC:1311099.1311750} mappings from C/C++ to Armv7-A and Armv8-A.}
  \begin{tabular}{|l|l|l|}
  \hline
  \textbf{Atomic Operation} & \textbf{Compiler Profile} & \textbf{Assembly Sequence}\\
  \hline
  \texttt{load(loc,sc)} & \texttt{clang -march=armv8 -O3} & 
   \cellcolor{armVIII}\texttt{LDA R0, [loc]}\\\cline{2-3}
  & \texttt{clang -march=armv7-a -O3} & 
  \cellcolor{armVII}\texttt{LDR R0, [loc]}\\
  & & \cellcolor{armVII}\texttt{DMB ISH}\\
  \hline
  \texttt{store(loc,val,sc)} & \texttt{clang -march=armv8 -O3} & 
   
    \cellcolor{armVIII}\texttt{MOV R1, \#val}\\
   & & \cellcolor{armVIII}\texttt{STL R1, [loc]}\\\cline{2-3}
  & \texttt{clang -march=armv7-a -O3} &
  
    \cellcolor{armVII}\texttt{MOV R1, \#val}\\
    & & \cellcolor{armVII}\texttt{DMB ISH}\\
    & & \cellcolor{armVII}\texttt{STR R1, [loc]}\\
    & & \cellcolor{armVII}\texttt{DMB ISH}\\
  \hline
  \end{tabular}
  \label{examplemappings}
\end{table*}

We combine instruction sequences into compiled litmus tests using a \textit{combining function} (Def.~\ref{comb}). Combining the compiled sequences produces exponentially many assembly litmus tests, all of which are valid combinations of the compiler profiles under test. We discuss how to prune this space of possible litmus tests, to prioritise litmus tests that are likely to be interesting, in \textsection\ref{exponent}.

\newpage\begin{definition}{\textit{Combining function}}.\label{comb} For a source litmus test $s$, and a set of compiled assembly sequences $\mathit{asms}$ produced by splitting and compiling the instructions of $s$:\\

\begin{tabular}{l l}
$\quad\mathit{combine}(\mathit{asms},\mathit{s}) = $ &
  $\mathit{let}\, mk\, (\mathit{asm}) = (\mathtt{init}=\mathit{s.\mathtt{init}},\; \mathtt{prog}=\mathit{asm},\; \mathtt{pred}=\mathtt{s.pred})\, \mathit{in}$ \\
  & $\mathit{map}\; mk\; \{\; \{\; \mathtt{instrs}=\mathit{mk\_asm}(\mathit{asms},\mathit{thread}),\, \mathtt{tid}=\mathit{thread}.\texttt{tid}\;\}\;$\\
  & $\quad\quad\quad\quad |\; \mathit{thread} \in \mathit{s.\mathtt{prog}}\; \}$\\
\end{tabular}\\

\begin{tabular}{l l}
$\;\;\;\,\mathit{mk\_asm}(\mathit{asms}, \mathit{thread}) = $ & $ \{\; \mathit{asm}\; |\; \mathit{src} \in \mathit{thread}.\mathtt{instrs},\, asm \in \mathit{asms},\,$\\
& $\mathit{where}\, \mathit{src}.\mathtt{iid} = asm.\mathtt{iid}\; \}$
\end{tabular}
\end{definition}

Each source test compiles to multiple assembly tests. We check if any \textit{c} in the set of compiled litmus tests \textit{C} exhibits a bug with respect to the source litmus test \textit{s}. We thus define a \textit{mixing bug}:

\begin{definition}{\textit{Mixing bug}}\label{mbug}: For a well-defined concurrent source program $s$ and its set $C$ of compiled litmus tests (given a splitting function, compiler profiles, and combining function):\\

$\mathit{MixingBug}(s,C) = \exists c \in C, \mathit{ConcurrencyBug}(s,c) \hspace*{\fill}(\textnormal{applies Def.~\ref{cfault}})$
\end{definition}

\begin{example}
Mix testing the test in Fig.~\ref{fig:SBexample}(a) produces Fig.~\ref{fig:SBexample}(d). Fig.~\ref{fig:SBexample}(d) arises when the operations of Fig.~\ref{fig:SBexample}(a) are compiled for Armv8-A and Armv7-A. Running Fig.~\ref{fig:SBexample}(d) under the Armv8 model produces the outcomes below. The mixing bug occurs since the load of \textit{y} on \texttt{P0} is compiled to the \texttt{``LDR;DMB''} sequence. Since the \texttt{LDR} instruction is missing a leading \texttt{DMB} barrier and it has no ordering semantics with respect to \texttt{STL}, it can reorder before the \texttt{STL} instruction on \texttt{P0}, leading to the outcome \texttt{\{P0:R0=0; P1:R0=0\}}.\\

\begin{center}
  \begin{tabular}{c}
  $\mathcal{B}(\texttt{P0} \Vert \texttt{P1},\mathcal{M}_{\mathit{Armv8}})$\\
  $\Downarrow$\\
\begin{lstlisting}[language=C, style=mystyle]
!!{ P0:R0=0; P1:R0=0; }!!
  { P0:R0=0; P1:R0=1; }
  { P0:R0=1; P1:R0=0; }
  { P0:R0=1; P1:R0=1; }
\end{lstlisting}\\
  \\
  Predicate satisfied---bug \xmark
  \end{tabular}
\end{center}
\end{example}

\subsection{Mix test notation}

Since mix testing generates many litmus tests we introduce a \textit{MixTest} notation in Fig.~\ref{symbols} to represent compiled tests that induce mixing bugs. A mix test is a labelled record consisting of a source litmus test (\texttt{test}) that is partitioned into its instructions and an \textit{assignment} function from profiles to the instructions they compile (\texttt{assignment}). A thread is split into its instructions, then each instruction (represented by its \texttt{IID}) is compiled using a profile $\mathit{comp}$ and sequenced together (\texttt{;}) to form a whole thread. The compiled test is represented using the mix test notation, for instance Fig.~\ref{fig:SBexample}(d) is $(\mathit{comp}_1(\texttt{P0\_0});\mathit{comp}_2(\texttt{P0\_1})) \Vert (\mathit{comp}_{1}(\texttt{P1\_0});\mathit{comp}_{2}(\texttt{P1\_1}))$ and the record in Fig.~\ref{notate}. 

\begin{figure}[H]
  \begin{tabular}{l}
  $\mathit{MixTest}\,= \{ \mathtt{test:}\, \mathit{LitmusTest}_{\mathit{src}}$,\\
  $\quad\quad\quad\quad\quad\mathtt{assignment:}\,  \mathit{CompilerProfile} \rightarrow \mathit{Set(Instructions)} \}$
  \end{tabular}
  \caption{Mix test notation.}
  \label{symbols}
\end{figure}

\begin{figure}[h]
\begin{tabular}{l r}
  \multicolumn{2}{c}{Fig.~\ref{fig:SBexample}(d) test = $(\mathit{comp}_1(\texttt{P0\_0});\mathit{comp}_2(\texttt{P0\_1})) \Vert (\mathit{comp}_{1}(\texttt{P1\_0});\mathit{comp}_{2}(\texttt{P1\_1}))$}.\\
  \\
  \begin{tabular}{l}
    Example:\\
    $\quad$ \{ \texttt{test} = Fig.~\ref{fig:SBexample}(a),\\
    $\quad\;\;$ \texttt{assignment} = \{\\
    \texttt{                    }$\quad\quad\quad\mathit{comp}_1\mapsto\{\texttt{P0\_0, P1\_0}\}$,\\
    \texttt{                    }$\quad\quad\quad\mathit{comp}_2\mapsto\{$\texttt{P0\_1,P1\_1}\}\}\}\\
    where:\\
    $\quad \mathit{comp}_{1}=\textnormal{\texttt{clang -march=armv8-a -O3}}$\\

    $\quad \mathit{comp}_{2}=\textnormal{\texttt{clang -march=armv7-a -O3}}$\\
  \end{tabular} &
  \begin{tabular}{l}
      \texttt{P0\_0 = store(x,1,sc)}\\
      \texttt{P0\_1 = load(y)}\\
      \texttt{P1\_0 = store(y,1,sc)}\\
      \texttt{P1\_1 = load(x)}\\\\
      $\mathit{comp}_1$(\texttt{P0\_0})\texttt{ = ``MOV;STL''}\\
      $\mathit{comp}_2$(\texttt{P0\_1})\texttt{ = ``LDR;DMB''}\\
      $\mathit{comp}_1$(\texttt{P1\_0})\texttt{ = ``MOV;STL''}\\ 
      $\mathit{comp}_2$(\texttt{P1\_1})\texttt{ = ``LDR;DMB''}\\    
  \end{tabular} 
\end{tabular}
\caption{Example of MixTest notation.}
\label{notate}
\end{figure}

\subsection{The choice of splitting function}\label{subsume}
The splitting function determines the set of instructions $I$ to be compiled separately. There is a trade-off here: the finer-grained the split, the more opportunities there are for problematic interactions between compiler mappings, but the larger the search space of possible sequences. The simplest function does not split the source test at all and just tests compilation using each profile $p \in P$, that is $|I|=1$. This corresponds to (non-mix) testing as conducted by prior work~\cite{10.1145/3460319.3469079,GeesonSmithCGO24,cmmtest,Chakraborty}. Mix testing strictly generalizes prior work, which compiles the whole program under one profile. The next function splits each source test $s$ into its constituent $K$ threads and compile those under different profiles, then $|I|=|K|$, and $|P|^{|K|}$ different choices. Intuitively bugs arise due to thread-local reordering~\cite{10.1145/1150019.1136489,cmmtest}, so if each thread is compiled using one mapping and each mapping is \textit{self}-consistent, then no bugs should arise. Since individual atomics mappings of each compiler have been rigorously tested, we expect most bugs found by splitting at the thread boundary are caught by non-mix testing. Therefore, we split litmus tests at the instruction level as shown in Fig.~\ref{splitting} ($|I|=4$ in this case), so that $I$ is bounded by the number of instructions of the input test. This function offers a good trade-off between the likelihood of finding bugs and the complexity of splitting the test into smaller fragments (given the simplicity of typical litmus tests).

\subsection{Putting it all together}
\begin{figure}[t]
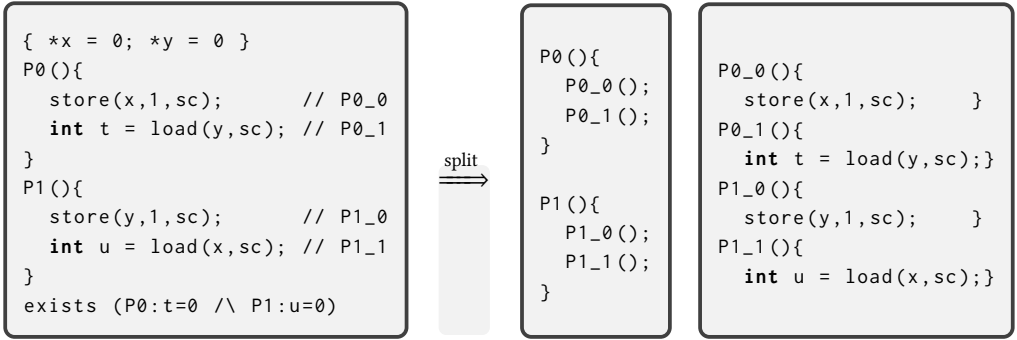

  \begin{tabular}{l l l l}
  \begin{tcolorbox}[hbox,left=2mm,right=2mm,top=2mm,bottom=2mm,boxsep=0mm]
\begin{lstlisting}[language=C, style=mystyle]
{ *x = 0; *y = 0 }
P0(){
  store(x,1,sc);      // P0_0 
  int t = load(y,sc); // P0_1 
}
P1(){
  store(y,1,sc);      // P1_0
  int u = load(x,sc); // P1_1
}
exists (P0:t=0 /\ P1:u=0)
\end{lstlisting}
  \end{tcolorbox} & 
  \begin{tcolorbox}[hbox,colframe=white,standard jigsaw,opacityback=0,bottom=21mm,,boxsep=-4mm]
  $\xRightarrow[]{\text{split}}$ 
  \end{tcolorbox}&
  \begin{tcolorbox}[hbox,left=2mm,right=2mm,top=3.8mm,bottom=3.8mm,boxsep=0mm]
\begin{lstlisting}[language=C, style=mystyle]
P0(){
  P0_0(); 
  P0_1();
}

P1(){
  P1_0();     
  P1_1();
}
\end{lstlisting}
  \end{tcolorbox} &
  \begin{tcolorbox}[hbox,left=2mm,right=2mm,top=5.8mm,bottom=5.8mm,boxsep=0mm]
\begin{lstlisting}[language=C, style=mystyle]
P0_0(){
  store(x,1,sc);    }  
P0_1(){
  int t = load(y,sc);}
P1_0(){
  store(y,1,sc);    } 
P1_1(){
  int u = load(x,sc);}
\end{lstlisting}
  \end{tcolorbox}\\
  \end{tabular}
  \caption{Splitting Fig.~\ref{fig:SBexample}(a) at the statement level produces multiple program instructions.}
  \label{splitting}
\end{figure}

\begin{figure}[H]
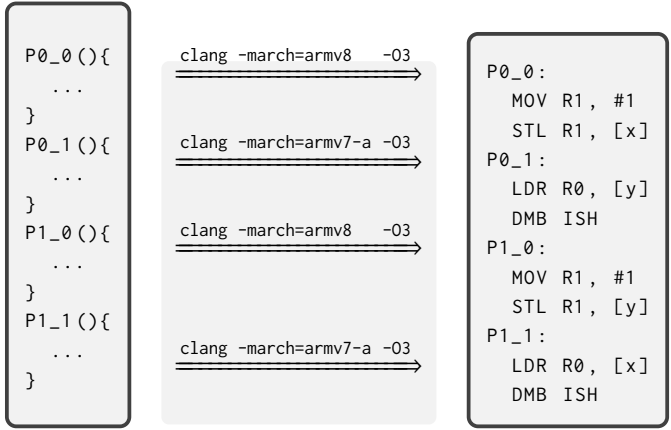

  \begin{tabular}{l l l}
  \begin{tcolorbox}[hbox,left=2mm,right=2mm,top=4mm,bottom=4mm,boxsep=0mm]
\begin{lstlisting}[language=C, style=mystyle]
P0_0(){
  ...
}
P0_1(){
  ...
}
P1_0(){
  ...    
}
P1_1(){
  ...
}
\end{lstlisting} 
  \end{tcolorbox}&
  \begin{tcolorbox}[hbox,colframe=white,standard jigsaw,opacityback=0,bottom=4mm,boxsep=-4mm]
  \begin{tabular}{l}
  $\xRightarrow[]{\texttt{clang -march=armv8\:\:\:\,\, -O3}}$
  \\\\
  $\xRightarrow[]{\vphantom{clang -march=armv7-a -O3}\texttt{clang -march=armv7-a -O3}}$
  \\\\
  $\xRightarrow[]{\texttt{clang -march=armv8\:\:\:\,\, -O3}}$\\\\
  \\
 $\xRightarrow[]{\texttt{clang -march=armv7-a -O3}}$\\\\

  \end{tabular}\end{tcolorbox} &
  \begin{tcolorbox}[hbox,left=2mm,right=2mm,top=2mm,bottom=2mm,boxsep=0mm]
\begin{lstlisting}[language=C, style=mystyle]
P0_0:
  MOV R1, #1
  STL R1, [x]
P0_1:
  LDR R0, [y]
  DMB ISH
P1_0:
  MOV R1, #1
  STL R1, [y]
P1_1:
  LDR R0, [x]
  DMB ISH
\end{lstlisting}
  \end{tcolorbox}
  \end{tabular}
  \caption{Compilation using multiple profiles produces compiled instruction sequences.}
  \label{compiling}
\end{figure}

\begin{figure}[b]
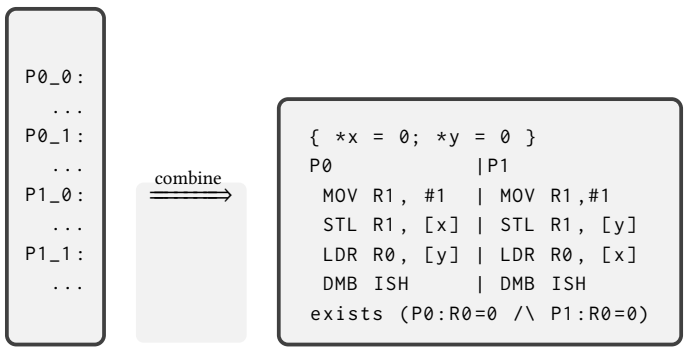

  \begin{tabular}{l l l}
  \begin{tcolorbox}[hbox,left=2mm,right=2mm,top=6mm,bottom=6mm,boxsep=0mm]
\begin{lstlisting}[language=C, style=mystyle]
P0_0: 
  ...
P0_1: 
  ...
P1_0: 
  ...
P1_1: 
  ...
\end{lstlisting} 
  \end{tcolorbox} &
  \begin{tcolorbox}[hbox,colframe=white,standard jigsaw,opacityback=0,bottom=20mm,boxsep=-4mm]
  \begin{tabular}{l}
  $\xRightarrow[]{\text{combine}}$
  \end{tabular} \end{tcolorbox}&
  \begin{tcolorbox}[hbox,left=2mm,right=2mm,top=2mm,bottom=2mm,boxsep=0mm]
  \begin{tabular}{c}
\begin{lstlisting}[language=C, style=mystyle]
{ *x = 0; *y = 0 }
P0           |P1
 MOV R1, #1  | MOV R1,#1
 STL R1, [x] | STL R1, [y]
 LDR R0, [y] | LDR R0, [x]
 DMB ISH     | DMB ISH            
exists (P0:R0=0 /\ P1:R0=0)
\end{lstlisting}
  \end{tabular}
  \end{tcolorbox}
  \end{tabular}
  \caption{Combining code and copying the initial state and predicate from Fig.~\ref{fig:SBexample} produces mixed tests.}
  \label{combining}
\end{figure}

Applying \tttool{} to Fig.~\ref{fig:SBexample}(a) is illustrated in Fig.~\ref{splitting}, Fig.~\ref{compiling}, and Fig.~\ref{combining}. \ttTool{} produces the mix test in Fig.~\ref{symbols} that represents Fig.~\ref{fig:SBexample}(d) using $(\mathit{comp}_1(\texttt{P0\_0});\mathit{comp}_2(\texttt{P0\_1})) \Vert$ $(\mathit{comp}_{1}(\texttt{P1\_0});$ $\mathit{comp}_{2}(\texttt{P1\_1}))$.

The litmus tests generated are simple tests where each instruction is a load, store, barrier, loop, or conditional operations. We parse the program inside the C/C++ litmus test and replace each instruction (\texttt{A},\texttt{B}) in the sequence \texttt{A;B} with \texttt{$comp_1$(A);$comp_2$(B)} (where $\mathit{comp_{1/2}}$ are profiles assigned by permutting all profiles under test). To handle conditionals or loops we check the condition for similar loads/stores, and then recurse into the loop body.

We use linkers to combine object files. We note that the linker will not apply link-time optimisation (LTO) unless the compilers used in the compilation and combination steps are the same. This is because LTO relies on GIMPLE and LLVM IR that is attached to object files, which is used to guide LTO. If the attached IRs of each file differ then the linker cannot optimise the assembly. When IRs differ the linker will instead emit branches to linked assembly code (instead of applying LTO to that code).

\begin{figure}[H]
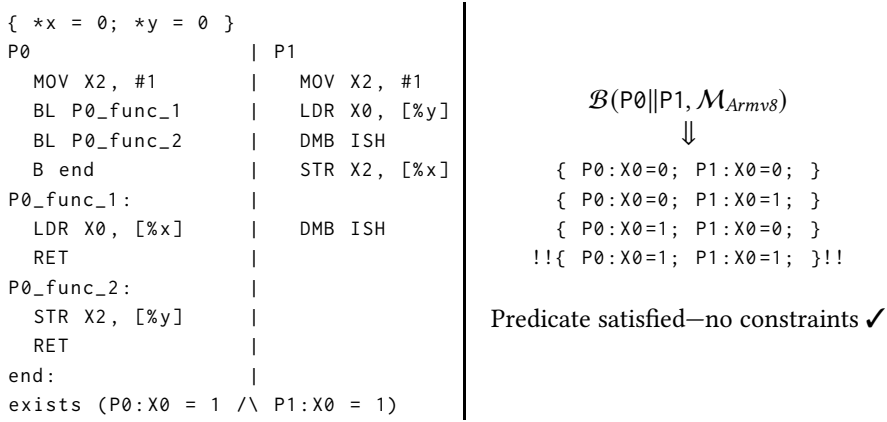

  \begin{tabular}{c | c}
\begin{lstlisting}[language=C, style=mystyle]
{ *x = 0; *y = 0 }
P0                 | P1       
  MOV X2, #1       |   MOV X2, #1     
  BL P0_func_1     |   LDR X0, [%y]
  BL P0_func_2     |   DMB ISH     
  B end            |   STR X2, [%x]
P0_func_1:         |
  LDR X0, [%x]     |   DMB ISH     
  RET              | 
P0_func_2:         |
  STR X2, [%y]     | 
  RET              | 
end:               | 
exists (P0:X0 = 1 /\ P1:X0 = 1)
\end{lstlisting} &
  \begin{tabular}{c}
  $\mathcal{B}(\texttt{P0} \Vert \texttt{P1},\mathcal{M}_{\mathit{Armv8}})$\\
  $\Downarrow$\\
\begin{lstlisting}[language=C, style=mystyle]
  { P0:X0=0; P1:X0=0; }
  { P0:X0=0; P1:X0=1; }
  { P0:X0=1; P1:X0=0; }
!!{ P0:X0=1; P1:X0=1; }!!
\end{lstlisting}\\
  \\
  Predicate satisfied---no constraints \cmark
  \end{tabular}
  \end{tabular}
  \caption{(Left) AArch64 Load Buffering test where C/C++ relaxed loads are compiled to branch instructions on \texttt{P0}. (Right) outcomes under the AArch64 model~\cite{aarch64}. The model allows the outcome \texttt{\{P0:X0=1; P1:X0=1\}}.}
  \label{blret}
\end{figure}

\subsection{The branching problem}\label{branching}
Splitting a test introduces function calls into each thread. A compiled litmus test has a corresponding branch instruction to the sequence that is separately compiled. For instance, GCC and LLVM generate \textit{branch-with-link} and \textit{return} instructions when targeting Armv8 AArch64. This can be problematic if processors implement the branch using a control-flow dependency that constrains the order of execution on each thread. Since we are looking for bugs that are exhibited by the re-ordering of observable events, we do not want to introduce such constraints.

Fortunately, each architecture we tested allows re-ordering across unconditional branches (Fig.~\ref{blret}). This means the effects of instructions after the branch can reorder before events of instructions prior. Intuitively, an unconditional branch is always taken, and so the micro-architecture is free to fill the pipeline with instructions on the branch taken as if the branch instruction itself was no-op. We empirically validated each compiler to check reordering (Fig.~\ref{blret}), and found both LLVM and GCC use call and return branches that allow reordering.

\subsection{The complexity of mix test generation}\label{exponent}

The number of compiled litmus tests we generate is exponential in the number of compiler profiles and number of program instructions. Mix testing takes a set of $S$ source litmus tests as input. Each $s \in S$ is split into a set of $I$ program instructions using a splitting function (Def.~\ref{strat}). Each $i \in I$ is compiled separately using each compiler profile $p$ in the set of $P$ compiler profiles. Each source litmus test then yields a set $C$ of different compiled litmus tests, where $|C|=|P|^{|I|}$. For example, \mixtesting{} Fig.~\ref{fig:SBexample} (a) yields $|P|=2$, $|I|=4$. That is $|C|=16$ possible compiled tests. The number of $|C|$ tests for each $s \in S$ rapidly increases as the size of the input test increases. We therefore reduce $S$, $P$ and $I$ as much as possible whilst maximising the coverage of code generation. We do so by:

\begin{description}
\item[\textbf{Curation of $P$}:] We omit compiler profiles that do not change the code generation of atomics relative to others. For instance \texttt{clang -O1,-O2,} and \texttt{-O3} use the same atomics mappings, but apply different optimisations. To maximise the chances of catching bugs we use \texttt{-O3}. We have worked with Arm's compiler experts to pick profiles that use different atomic mappings targeting Arm assembly. Depending on experts is a limitation (\textsection\ref{limits}) we accept, but are open to automated techniques that find mappings as they arise.

\item[\textbf{Symmetry reduction on $S$}:] We do not generate source tests where the contents of each thread are simply swapped.

\item[\textbf{Bound $|I|$ by fixing the splitting function}:] The number of source instructions is determined by the splitting function. By splitting litmus tests at the instruction level we bound the number of generated tests, although the exponential complexity remains in theory. The number of compiled tests $|C|$ for each $s \in S$ is still exponential in $I$ and $P$, and it is possible that duplicate tests exist in each set $C$. In the worst case when all compiler mappings in $P$ are disjoint, that is a given C/C++ atomic operation compiles to a different instruction for each profile $p \in P$, the complexity is $|P|^{|I|}$ for each $s \in S$. In the best case when every compiler implements one set of mappings, we only need to test one compiler, and so $|P|=1$ and the complexity is $1^{|I|}$ or $1$ for each $s \in S$. The best case rarely happens in practice, since a given architecture has multiple possible atomics mappings, for each architecture sub-version, and hence multiple compiler profiles to test. For example, LLVM implements atomics differently for at least Armv8, Armv8.1, Armv8.2, Armv8.3, and Armv8.4. Further, new atomic instructions are announced with new architecture versions to improve performance of concurrent workloads. This means \mixtesting{} the compilation of concurrent programs is unfortunately a practical necessity at least until everyone agrees on an ABI that specifies common atomics mappings. Lastly, compilers are routinely revised and the code they generate often changes. As such our analysis is not exhaustive or even timeless, and we must periodically revise $P$ and $S$ as compilers are updated. It is not however surprising that the compiler profiles and tests suites must be updated.
\end{description}

\subsection{The scope of our testing}\label{do}

\begin{description}

\item[\textbf{ISA-compatible assembly}:] We mix test compiled programs that are instruction-set architecture (ISA) \textit{compatible}. This means their binary representation can be combined and executed without fault, as permitted by the envelope of the ISA. We do not yet require \textit{atomics ABI-compatibility} in as far as we cannot find any official atomics ABIs outside of what we contribute in \textsection\ref{imp}, but we do require that compiled programs are ABI-compatible in every other way (procedure call standards, exception handling, and so on). In general, mix-testing applies to code generated for CPUs of any architecture that can co-exist in the same shared-memory system, but for this work we limit our focus to a subset of recent Arm architectures.

\item[\textbf{Redundant mappings}:] It is desirable to omit repeated testing of instructions whose implementation remains the same. For instance, both LLVM and GCC implement a C/C++ \texttt{atomic\_fence} operation as a \texttt{DMB} memory barrier when targeting the Armv8 architecture. In this case compiling using only one profile \textit{should} suffice. Unfortunately, compilers do not implement common mappings. For instance, Arm's partners highlight~\cite{john} that MSVC emits two barriers for atomic read-modify-write operations, whereas LLVM emitted one at that time. Omitting test of mappings without an official ABI is risky and can lead to missed bugs. We therefore check redundant mappings and develop an Armv8 atomics ABI to define the envelope of compiler conformance going forward.
\end{description}

\newpage\section{The \ttTool{} tool implementation}\label{impl}

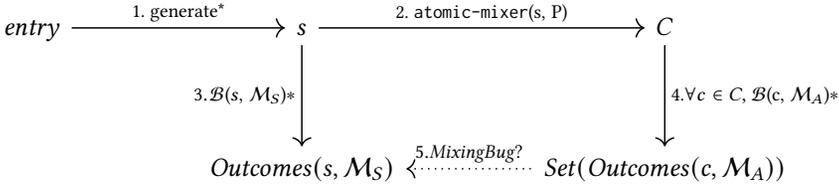
\begin{figure}[h]
\centering
  \begin{tikzcd}[ampersand replacement=\&, row sep=huge, column sep=huge]
    \mathit{entry} \arrow[r, "\text{1. generate*}"]
    \& s
      \arrow[r, "\text{2. \tttool{}(s, P)}"]
      \arrow[d, swap, "3. \text{$\mathcal{B}$($s$, $\mathcal{M}_{S}$)}*"]
    \& C
      \arrow[d, "\text{4.$\forall c \in C,\mathcal{B}$(c, $\mathcal{M}_{A}$)}*"]\\
    \& \mathit{Outcomes}(s,\mathcal{M}_{S})
      \arrow[r, leftarrow, dotted, "5.\mathit{MixingBug}?"]
    \& \mathit{Set(Outcomes}(c,\mathcal{M}_{A}))
    \end{tikzcd}
  \caption{Mix testing technique implementation using the new \tttool{} tool and prior work*~\cite{GeesonSmithCGO24,herdtools}.}\label{tele}
\end{figure}

\subsection{Technique and tool implementation}\label{tool}
We present the \tttool{} tool and mix testing technique implementation. Fig.~\ref{tele} shows how we implement the \mixtesting{} technique. We describe the \mixtesting{} process as follows:

\begin{enumerate}
\item Generate a concurrent C/C++ litmus test $s$.
\item Given a set $P$ of compiler profiles, apply \texttt{\tttool{}(s,P)} to get a set $C$ of compiled tests.
\item Collect $\mathit{Outcomes}(s,\mathcal{M}_{S})$: the outcomes of simulating $s$ under the source model $\mathcal{M}_{S}$ (Def.~\ref{out}).
\item For each $c \in C$ collect $\mathit{Outcomes}(c,\mathcal{M}_{A})$: the outcomes under its architecture model $\mathcal{M}_{A}$.
\item Check for mixing bugs (Def.~\ref{mbug}). 
\end{enumerate}

We use the \texttt{Memalloy}~\cite{Wickerson:2017:ACM:3009837.3009838} and \texttt{diy}~\cite{Alglave:2012:FWM:2205506.2205523} litmus test generators to produce tests $S$. We simulate source and compiled tests using the \texttt{herd}~\cite{herdtools} simulator. We compare program outcomes using the \texttt{mcompare} tool~\cite{herdtools}. The \tttool{} tool itself extends the \telechat{} toolchain~\cite{GeesonSmithCGO24}, which handles the non-mix testing case by generating one compiled litmus test for each profile. The \tttool{} tool increases coverage by enumerating atomics mappings of multiple profiles. Provided prior work is kept up to date, \tttool{} is future-proof against the future evolution of programming languages, architectures, and their underlying memory models.

By extending \telechat{}, \tttool{} inherits a deterministic framework under authoritative models. Prior testing tools~\cite{cmmtest,10.1145/3460319.3469079} execute compiled programs on hardware to collect program outcomes (Def.~\ref{out}). Hardware vendors are not however required to implement all behaviour permitted by the architecture specification. Consequently hardware-based testing may not exhibit all program behaviours, and bugs. Geeson and Smith~\cite{GeesonSmithCGO24} address this issue by parametrising testing under executable source and architecture models of Armv8 AArch64~\cite{aarch64} (official), RISC-V~\cite{lucrisc} (official), RC11~\cite{simonrc11}, Armv7 (unofficial)~\cite{armv7}, Intel x86-64~\cite{x86tso}, MIPS~\cite{mips}, IBM PowerPC~\cite{ppc}, and more. By using models and simulation, \tttool{} deterministically covers all possible outcomes of each test that terminates (up to bounds on loop unrolling).

\subsection{Challenges faced during implementation}\label{challenge}

The key to efficient \mixtesting{} is knowing the number of mappings of a given atomic operation. There are many different compilers (for example, GCC and LLVM) and their code generation may change at any time to support new architectures, new optimisations, or modifications of existing implementations. A naïve approach is to test all compiler releases for each architecture. Despite the theoretical possibility that each release implements entirely different code generation, the reality is that few changes to atomics occur in practice. We therefore look for changes in code generation and only test each variant once. We describe simulation penalty and how we address it.

\begin{description}

\newpage\item[\textbf{Simulation penalty}:] Simulating the behaviour of litmus tests under models is computationally complex. We must simulate each source program $s \in S$ to collect its behaviours. Then, for each such $s$ we must simulate for every $c \in C$, where $C$ is the set of compiled programs derived from $s$ by \tttool{}. Our goal is thus to reduce the number and size of the target test sets $C$ for each $s$ whilst increasing the coverage of code generated by compilers. We do so by hashing the generated assembly code of the litmus test.

We group $C$ by hashes and check one representative of each group. It is possible that changing the compiler profile only changes one or two atomics mappings whilst other mappings remain unchanged. For example Armv8.3-A changes the mapping of acquire loads to use the \texttt{LDAPR} instruction instead of the existing Armv8 \texttt{LDAR} instruction. Since all other atomics remain unchanged many compiled programs will have the same static hash and behaviour under simulation. Only one program with a given hash needs to be simulated in a group. We compute hashes using the \texttt{mshowhashes} tool~\cite{herdtools}. By doing so we only need to simulate one test from each group of tests with the same hash. Hashing drastically reduces the number of tests we must consider.

\item[\textbf{Handling programs that are larger than litmus tests}:] We limit our focus to small litmus tests of up to five threads, and up to 20 lines of code. The bottleneck of our flow is the \texttt{herd} simulator, which we use to compute the allowed behaviours of source and assembly tests under source and target models, respectively. As the number of threads grows, exhaustive simulation quickly becomes infeasible. It may be possible to use other tools, but \texttt{herd} is attractive for our current needs since it is easily extensible. That said, we don't expect much would be gained by moving to larger test cases, since we are focusing on testing mappings interoperability, and bugs that cannot be expressed using small litmus tests are rare (but certainly not impossible -- we are aware of one bug~\cite{sunghwan}).

\item[\textbf{Mix testing is I/O bound}:] Even if we can discard duplicates using hashing, \tttool{} must still \textit{generate} them. \tttool{} must generate all $c \in C$ as \texttt{mshowhashes} cannot compute the hashes until it has tests. 

\item[\textbf{Adapting to changing architectures and language standards}:] This is a challenge for any concurrency testing tool. The C/C++ language and Arm Architecture specifications undergo numerous changes as implementors provide feedback and new requirements. Memory models, and testing tools that rely on models, must adapt in turn. We use \texttt{herd} to simulate the source C program and the compiled program. \texttt{herd} takes a cat~\cite{herdtools} file describing the desired model, so changing language is a simple matter of changing this file. The models changed several times during this project, causing no issues.
\end{description}

\section{Evaluation}\label{eval}
We evaluated the \mixtesting{} technique and \tttool{} tool by conducting a number of case studies using LLVM and GCC. We show that mix testing strictly generalises testing with respect to a single compiler profile by using \tttool{} to find (non mixing) bugs that prior work is limited to being able to find (\textsection\ref{repro}), and discover four previously unknown mixing bugs (\textsection\ref{missed}) of which one was found manually (\textsection\ref{struct2}). We also found a mixing bug in mappings proposed for the JVM (\textsection\ref{java}). We cover the limitations of \mixtesting{} (in \textsection\ref{limits}).

\subsection{Reproducing an existing (non-mixing) bug}\label{repro}

Since \mixtesting{} with one profile corresponds to non-mix testing it follows that \tttool{} should be able to reproduce existing bugs. We reproduce a (non-mixing) bug found by Geeson and Smith 2024~\cite{GeesonSmithCGO24} using \tttool{}.

\begin{example}
Consider the Message Passing test in Fig.~\ref{mp_rc11} (left). Fig.~\ref{mp_rc11} (middle) shows that the outcome \texttt{\{P1:r0=0; y=2\}} is forbidden by the RC11 model~\cite{simonrc11}. When compiled to target Armv8.2-A the compiled program (Fig.~\ref{swpbug}) exhibits the outcome under the Armv8 AArch64~\cite{aarch64} model. Fig.~\ref{mp_rc11} (right) shows the outcomes of \tttool{}-generated test (Fig.~\ref{swpbug}) allowed by the Armv8 AArch64 model~\cite{aarch64}. These outcomes match those in the bug report~\cite{swp}. This bug has since been fixed in LLVM by Arm's engineers.

\begin{figure}[h]
\centering
  \begin{tabular}{c | c}
\begin{lstlisting}[language=C, style=mystyle]
{ *x = 0; *y = 0; }
P0() {
  store(x,1,rlx);
  fence(rel);
  store(y,1,rlx);
}
P1() {
  atomic_exchange(y,2,rel);
  fence(acq);
  int r0 = load(x,rlx);
}
exists (P1:r0=0 /\ y=2)
\end{lstlisting} &
  \begin{tabular}{c}
  \begin{tabular}{c | c}
  $\mathcal{B}(\texttt{P0} \Vert \texttt{P1},\mathcal{M}_{\mathit{C/C++}})$
  & $\mathcal{B}(\mathit{comp}(\texttt{P0}) \Vert \mathit{comp}(\texttt{P1}),\mathcal{M}_{\mathit{Armv8}})$\\
  $\Downarrow$ & $\Downarrow$\\
\begin{lstlisting}[language=C, style=mystyle]
{ P1:r0=0; y=1; }
	
{ P1:r0=1; y=1; }
{ P1:r0=1; y=2; }
\end{lstlisting} &
\begin{lstlisting}[language=C, style=mystyle]
  { P1:r0=0; y=1; }
!!{ P1:r0=0; y=2; }!!
  { P1:r0=1; y=1; }
  { P1:r0=1; y=2; }
\end{lstlisting}\\
  $\Downarrow$ & $\Downarrow$\\
  Predicate not satisfied \cmark & Predicate satisfied---bug \xmark\\
  \end{tabular}\\\\
  $\mathit{comp} = \texttt{"clang -march=armv8.2-a -O3"}$
  \end{tabular}
  \end{tabular}
  \caption{\tttool{} finds a non-mixing bug~\cite{swp}. rlx = \texttt{relaxed}, rel = \texttt{release}, and acq = \texttt{acquire}.}
  \label{mp_rc11}
\end{figure}
\end{example}

\subsection{Finding bugs the state-of-the-art cannot}\label{missed}

Mix testing can find bugs that current tools cannot, since they require mixing and are thus out of scope. We checked a compiler patch~\cite{CASPAL} and found and reported~\cite{llvm-mix-bug-2} a mixing bug that was missed by Geeson and Smith when they tested it back in January 2023. This example highlights the difficulty of testing the compilation of concurrency as a problem that cannot be addressed by testing atomics mappings in isolation, but rather by strategic testing in the presence of exponentially many choices of mappings. Mix testing takes the field forward both in terms of what is possible conceptually (mixing bugs) and what is possible in today's tools.

\begin{figure}[h]
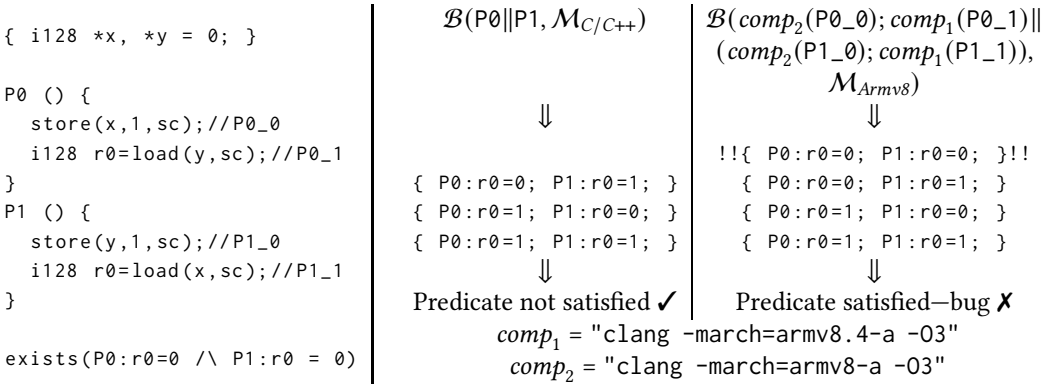

\centering
  \begin{tabular}{c | c}
\begin{lstlisting}[language=C, style=mystyle]
{ i128 *x, *y = 0; }

P0 () {
  store(x,1,sc);//P0_0
  i128 r0=load(y,sc);//P0_1
}
P1 () {
  store(y,1,sc);//P1_0
  i128 r0=load(x,sc);//P1_1
}

exists(P0:r0=0 /\ P1:r0 = 0)
\end{lstlisting} &
  \begin{tabular}{c}
  \begin{tabular}{c | c}
  $\mathcal{B}(\texttt{P0} \Vert \texttt{P1},\mathcal{M}_{\mathit{C/C++}})$
  &$\mathcal{B}(\mathit{comp}_{2}(\texttt{P0\_0});\mathit{comp}_{1}(\texttt{P0\_1})\Vert$\\
  &$(\mathit{comp}_{2}(\texttt{P1\_0});\mathit{comp}_{1}(\texttt{P1\_1})),$\\
  &$\mathcal{M}_{\mathit{Armv8}})$\\
  $\Downarrow$ & $\Downarrow$\\
\begin{lstlisting}[language=C, style=mystyle]
	
{ P0:r0=0; P1:r0=1; }
{ P0:r0=1; P1:r0=0; }
{ P0:r0=1; P1:r0=1; }
\end{lstlisting} &
\begin{lstlisting}[language=C, style=mystyle]
!!{ P0:r0=0; P1:r0=0; }!!
  { P0:r0=0; P1:r0=1; }
  { P0:r0=1; P1:r0=0; }
  { P0:r0=1; P1:r0=1; }
\end{lstlisting}\\
  $\Downarrow$ & $\Downarrow$\\
  Predicate not satisfied \cmark & Predicate satisfied---bug \xmark\\
  \end{tabular} \\
  $\mathit{comp}_1$ = \texttt{"clang -march=armv8.4-a -O3"}\\
  $\mathit{comp}_2$ = \texttt{"clang -march=armv8-a -O3"}
  \end{tabular}
  \end{tabular}
  \caption{Mixing bug~\cite{llvm-mix-bug-2}. The outcome \texttt{\{P0:r0=0; P1:r0=0\}} is forbidden by the C/C++ model~\cite{C11}, but the compiled program exhibits it under the AArch64 model~\cite{aarch64}. sc = \texttt{seq\_cst}, and i128 = \texttt{\_Atomic \_\_int128}.}
  \label{missed-bug}
\end{figure}

\begin{example}
Consider the test in Fig.~\ref{missed-bug} (left). Fig.~\ref{missed-bug} (middle) shows that the outcome of the exists clause \texttt{\{P0:r0=0; P1:r0=0\}} is forbidden by the C/C++ model~\cite{C11}. Fig.~\ref{missed-bug} (right) shows a mixing bug arises when mix testing the source test using profiles targeting Armv8-A and Armv8.4-A. Fig.~\ref{missed-bug-a64} shows the mix test generated by \tttool{}. In this case the load pair (\texttt{LDP}) of $x$ on \texttt{P1} has no leading barrier, and since \texttt{LDP} has no ordering semantics, its effects can be reordered before the store-release exclusive pair instruction (\texttt{STLXP}) on \texttt{P1}. The compiled program exhibits the outcome \texttt{\{P0:r0=0; P1:r0=0\}} under the AArch64 model~\cite{aarch64}.

\begin{figure}[h]
\centering
  \begin{tabular}{l |l}
\begin{lstlisting}[language=C, style=mystyle]
{ *x = 0; *y = 0; }
P0                |P1 
 MOV X2, #1       | MOV X6, #1                     
 DMB ISH          |loop:LDAXP X1,X2,[%P1_y]
 STP X1,X2,[%P0_x]| STLXP W4,X5,X6,[%P1_y]
 DMB ISH          | CBNZ W4, loop          
 LDP X4,X0,[%P0_y]| LDP X4, X0, [%P1_x]       
 DMB ISH          | DMB ISH         

exists (P0:X0 = 0 /\ P1:X0 = 0)
\end{lstlisting} &
  \begin{tabular}{l}
  $\mathit{MixTest}\, = \{$\\
  $\quad\quad\textnormal{\texttt{test}=Fig.~\ref{missed-bug} (left)},$\\
  $\quad\quad\texttt{assignment}=\textnormal{map}\}$\\
  where:\\
  map=\texttt{\{$\mathit{comp}_1\mapsto$\{P0\_0,P0\_1,P1\_1\},}\\
  $\quad\quad\,\,\,\,\,\mathit{comp}_2\mapsto$\texttt{\{P1\_0\} \}}\\
  $\mathit{comp}_{1}\textnormal{=\texttt{"clang -march=armv8.4-a -O3"}}$\\
  $\mathit{comp}_{2}\textnormal{=\texttt{"clang -march=armv8-a -O3"}}$
  \end{tabular}
  \end{tabular}
  \caption{Mix test that exposes mixing bug: $(\mathit{comp}_{1}(\texttt{P0\_0});\mathit{comp}_{1}(\texttt{P0\_1}))\Vert(\mathit{comp}_{2}(\texttt{P1\_0});\mathit{comp}_{1}(\texttt{P1\_1}))$.}
  \label{missed-bug-a64}
\end{figure}
\end{example}

Until now, only experts in both compilers and concurrency would be likely to find such a bug. The bug would not be caught by the state-of-the-art tools, since they do not conduct \mixtesting{}. The test in Fig.~\ref{missed-bug} is not unusual by concurrency standards, but the mixing bug is likely detectable only if the user has detailed knowledge of the atomics mappings in today's compilers and the concurrency experience needed to reproduce it. Indeed, neither concurrency architects nor engineers caught this bug. We worked closely with Arm's engineers to report the bug~\cite{llvm-mix-bug-2}. In the process, we developed \mixtesting{} and fostered a team of compiler engineers who handle queries regarding the compilation of atomics going forward\footnote{For compiler and ABI inquiries please contact: \url{arm.eabi@arm.com}}. 

We found three mixing bugs automatically~\cite{llvm-mix-bug,gcc-mix-bug,llvm-mix-bug-2}, and one bug manually~\cite{wilcostruct}. 

\begin{enumerate}
\item 32-bit sequentially consistent load is missing a barrier: See Fig.~\ref{fig:SBexample}, report~\cite{llvm-mix-bug}, and \textsection\ref{practice}.
\item 64-bit sequentially consistent load is missing a barrier: See report~\cite{gcc-mix-bug}. An analogue of (1), but for 64-bit loads when compiling to target 32-bit systems.
\item 128-bit sequentially consistent load is missing a barrier: See report~\cite{llvm-mix-bug-2}, Fig.~\ref{missed-bug}, and \textsection\ref{missed}.
\item \texttt{\_Atomic} struct size and alignment differ between LLVM and GCC. See report~\cite{wilcostruct}, Fig.~\ref{struct}, and \textsection\ref{struct2}.
\end{enumerate}

Each bug is triggered by a different tests and profiles. These tests were found using variants of store buffering tests with either sequentially consistent stores or read-modify-write operations. Picking the size of accesses from 32, 64, or 128-bits triggers different code paths in GCC and LLVM. Further, the Armv7 and Armv8 AArch64 back-ends are different targets in LLVM, and their code-generation is triggered by different compiler profiles. In other words we found three unique bugs. Generally, the test inputs and compiler profiles are not \textit{orthogonal}. The choice of test and profile cannot be arbitrarily varied but must be chosen to find bugs. This is problematic, since the search space is exponential in these inputs (\textsection\ref{exponent}). Mix testing thus relies on good choices of profiles and tests to trigger the conditions for bugs. 

It is reasonable to question whether mixing bugs only arise when mixing acquire-release and barrier-based implementations. We now explore a mixing bug that does not require barriers.

\subsection{Finding mixing bugs in proposed mappings}\label{java}
One of Arm's partners approached Arm's compiler teams with a proposal to the change the default mappings (Table.~\ref{ldapr}) of sequentially consistent~\cite{Lamport:1979:MMC:1311099.1311750} loads and stores when compiling for the release consistency processor consistency extension (RCPC). The RCPC extension introduces the \texttt{LDAPR} instruction whose effects can reorder before prior store-release (\texttt{STLR}) instructions that access different memory locations. The \texttt{LDAPR} instruction has the potential~\cite{kyrylo} to improve performance over the \texttt{LDAR} instruction (the current load implementation). Replacing \texttt{LDAR} with \texttt{LDAPR} alone however is unsound since it can reorder with prior stores (\texttt{STLR}). Instead, Arm's partners proposed to both strengthen the \texttt{STLR} with a trailing barrier (\texttt{DMB ISH}), and relax loads to use \texttt{LDAPR}, effectively preventing non-mixing bugs. We applied mix testing to show that this case would not be correct when mixing their proposed mappings in with code targeting Armv8-A.

\begin{table*}[h]
  \caption{One of Arm's partners asked if relaxing SC loads, and strengthening SC stores would be sound.}
  \begin{tabular}{|l|l|l|}
  \hline
  \textbf{Atomic Operation} & \textbf{Compiler Profile} & \textbf{Assembly Sequence} \\
  \hline
  \texttt{load(loc,sc)} & \texttt{clang -march=armv8 -O3 (current)} & 
   \texttt{LDAR W0, [loc]}\\\cline{2-3}
  & \texttt{clang -march=proposed -O3} &
  \texttt{LDAPR W0, [loc]}\\
  \hline
  \texttt{store(loc,val,sc)} & \texttt{clang -march=armv8 -O3 (current)} & 
  \texttt{MOV W1, \#val}\\
  & & \texttt{STLR W1, [loc]}\\\cline{2-3}
  & \texttt{clang -march=proposed -O3} &
  \texttt{MOV W1, \#val}\\
  & & \texttt{STLR W1, [loc]}\\
  & & \texttt{DMB ISH}\\
  \hline
  \end{tabular}
  \label{ldapr}
\end{table*}

\begin{example}
Consider the test in Fig.~\ref{ldaprbug} (left). Fig.~\ref{ldaprbug} (middle) shows that the outcome of the exists clause \texttt{\{P0:r0=0; P1:r0=0\}} is forbidden by the C/C++ model~\cite{C11}. Fig.~\ref{ldaprbug} (right) shows a mixing bug arises when mix testing the source test using the mappings in Table.~\ref{ldapr}. Fig.~\ref{ldaprcase} shows the mix test we manually found (no compiler implements the proposed mappings without which \tttool{} cannot work). In Fig.~\ref{ldaprcase} the store-release (\texttt{STLR}) instruction on \texttt{P0} has no trailing barrier, and the effects of executing the \texttt{LDAPR} can be reordered before the effects of the \texttt{STLR}. The compiled program exhibits the outcome \texttt{\{P0:r0=0; P1:r0=0\}} under the AArch64 model~\cite{aarch64}.

\begin{figure}[h]
\centering
  \begin{tabular}{c | c}
\begin{lstlisting}[language=C, style=mystyle]
{ *x = 0; *y = 0 }

P0 () {
  store(x,1,sc);//P0_0
  int r0=load(y,sc);//P0_1
}
P1 () {
  store(y,1,sc);//P1_0
  int r0=load(x,sc);//P1_1
}

exists (P0:r0=0 /\ P1:r0=0)
\end{lstlisting} &
  \begin{tabular}{c}
  \begin{tabular}{c | c}
  $\mathcal{B}(\texttt{P0} \Vert \texttt{P1},\mathcal{M}_{\mathit{C/C++}})$
  &$\mathcal{B}((\mathit{comp}_{1}(\texttt{P0\_0});\mathit{comp}_{2}(\texttt{P0\_1}))$\\
  &$\Vert (\mathit{comp}_{1}(\texttt{P1\_0});\mathit{comp}_{2}(\texttt{P1\_1})),$\\
  & $\mathcal{M}_{\mathit{Armv8}})$\\
  $\Downarrow$ & $\Downarrow$\\
\begin{lstlisting}[language=C, style=mystyle]
	
{ P0:r0=0; P1:r0=1; }
{ P0:r0=1; P1:r0=0; }
{ P0:r0=1; P1:r0=1; }
\end{lstlisting} &
\begin{lstlisting}[language=C, style=mystyle]
!!{ P0:r0=0; P1:r0=0; }!!
  { P0:r0=0; P1:r0=1; }
  { P0:r0=1; P1:r0=0; }
  { P0:r0=1; P1:r0=1; }
\end{lstlisting}\\
  $\Downarrow$ & $\Downarrow$\\
  Predicate not satisfied \cmark & Predicate satisfied---bug \xmark\\
  \end{tabular} \\
  $\mathit{comp}_1$ = \texttt{"clang -march=armv8 -O3"}\\
  $\mathit{comp}_2$=\texttt{"clang -march=proposed -O3"}
  \end{tabular}
  \\
  \end{tabular}
  \caption{A mixing bug arises if $\mathit{comp}_1$ and $\mathit{comp}_2$ mappings are mixed. The outcome \texttt{\{P0:r0=0; P1:r0=0\}} is forbidden by the C/C++ model~\cite{C11}, but the compiled program exhibits it under AArch64~\cite{aarch64}. sc = \texttt{seq\_cst}.}
  \label{ldaprbug}
\end{figure}
\end{example}

\begin{figure}[h]
\centering
  \begin{tabular}{l | l}
\begin{lstlisting}[language=C, style=mystyle]
{ *x = 0; *y = 0 }
 P0               | P1
  MOV W1, #1      |   MOV W1, #1
  STLR W1,[%P0_x] |   STLR W1,[%P0_y]
  LDAPR W0,[%P0_y]|   LDAPR W0,[%P0_x]

exists (P0:X0=0 /\ P1:X0=0)
\end{lstlisting} &
  \begin{tabular}{l}
  $\mathit{MixTest}\, = \{$\\
  $\quad\quad\quad \texttt{test}=\textnormal{Fig.~\ref{ldaprbug} (left)},$\\
  $\quad\quad\quad\texttt{assignment}=\textnormal{map}\}$\\
  where:\\
  map = \texttt{\{ $\mathit{comp}_1\mapsto$\{P0\_0,P1\_0\},}\\
  $\quad\quad\quad\,\,\,\,\,\mathit{comp}_2\mapsto$\texttt{\{P0\_1,P1\_1\} \}}\\
  $\mathit{comp}_{1}=\textnormal{above},\,\mathit{comp}_{2}=\textnormal{above}$
  \end{tabular}
  \end{tabular}
  \caption{Mixing bug without barriers: $(\mathit{comp}_{1}(\texttt{P0\_0});\mathit{comp}_{2}(\texttt{P0\_1}))\Vert(\mathit{comp}_{1}(\texttt{P1\_0});\mathit{comp}_{2}(\texttt{P1\_1}))$.}
  \label{ldaprcase}
\end{figure}

There is nothing wrong with the proposed mappings, provided \textit{all} stores are strengthened. In general we cannot know if a compilation unit will be mixed with other code for different (yet compatible) architectures. As long as multiple mappings exist, they may be mixed. The user can either guarantee the whole program is \textit{always} compiled using the proposed mappings or otherwise every compiler implementation must change. This requires that every compiler that supports Armv8-A and above (including LLVM, GCC, and MSVC) strengthens their SC stores with a trailing barrier. Unfortunately such a wide reaching change is unlikely to be accepted in practice. This proposal constitutes an ABI break with respect to today's compilers.

It is possible to use the proposed mappings without mixing bugs. Arm's partner wanted to change the Java Virtual Machine (JVM) implementation to use the proposed mappings. When used in isolation these mappings are sound, since the JVM uses a JIT compiler that can dynamically generate code using the proposed mappings all at once. However there are three cases where mixing bugs can arise. Firstly, heap locations may be written to by the JVM's C++ code using SC atomics (the \texttt{STLR} instruction), but then later read by Java volatiles (using \texttt{LDAPR}). This can be fixed by inserting barriers after every C/C++ store in the JVM source. Secondly, a user's C++ code may share a memory buffer with Java code (for example, a \texttt{java.nio.ByteBuffer}), where the C/C++ code stores to the buffer and Java loads from it (using \texttt{VarHandle::getVolatile}). Again, the user must insert barriers after C/C++ stores. Thirdly, bugs may arise if the JVM interacts with C/C++ through foreign function interfaces (FFI) such as JNI (for instance using an API call to \texttt{SetIntField}). Assuming the JVM does not synchronize at FFI boundaries (see \textsection\ref{branching}), barriers must added here too. Assuming these cases are handled, there are (probably) no mixing bugs.

\subsection{Mixing bugs in \texttt{\_Atomic struct} implementations}\label{struct2}

This bug~\cite{wilcostruct} was discovered manually while developing the ABI in \textsection\ref{imp}. Manual effort was required since \tttool{} depends on \texttt{herd}, which doesn't support \texttt{struct}s. This bug arises when two different compilers translate code for the same ISA. We discovered that GCC and LLVM have incompatible implementations of \texttt{\_Atomic struct}s. Both the size and alignment requirement calculated in Fig.~\ref{struct} differs between compilers. The \texttt{sizeof} operator is used to determine the storage allocation and size of atomic instructions to be used. In this case GCC's engineers chose to use an inefficient locking call (\texttt{atomic\_load}), whereas LLVM's engineers used a load acquire (\texttt{LDAR}) instruction. GCCs engineers chose to use the locking call, since there aren't any instructions to handle unaligned atomics or oddly-sized types, LLVMs engineers chose to use \texttt{LDAR} on the basis that every other access would share the same alignment values. Mixing code generated by both is problematic since LLVM may write struct padding bits to memory where GCC allocates entirely unrelated data---mixing LLVM and GCC code can invalidate data and hence program execution in unknown ways. The solution is to overalign and pad atomic types to the next supported atomic size.

\begin{figure}[t]
  \begin{tabular}{c}
\begin{lstlisting}[language=C, style=mystyle]
typedef _Atomic struct { char a[5]; } X;
int size_x = sizeof(X); // GCC=5, LLVM=8
int align_x = __alignof(X); // GCC= 1, LLVM=8

X f(X *p) { return *p; } 
// GCC=bl __atomic_load, LLVM=LDAR X2, [loc]

X g(X *p[2]) { return *p[1]; }
// GCC=__atomic_load, LLVM=LDR X1, [loc,#8]; LDAR X2, [X1]
\end{lstlisting}
  \end{tabular}
  \caption{Mixing struct implementations. This issue also effects x86 code generation.}
  \label{struct}
\end{figure}

\subsection{Limitations}\label{limits}
There are three limitations that contribute to the complexity of \mixtesting{}. We rely on experts to provide the compiler profiles, we depend on model-based tooling, and require test generators that have good atomics coverage.

Fortunately, there are practical solutions to these issues. Typically, only a small number of compiler profiles introduce \textit{new} atomics mappings, and so we only need to test those. Second, we assume herdtools~\cite{herdtools} implements all instructions we test. We thus added new features to \texttt{herd} including an 128-bit signed integer type~\cite{i128,i1282} to handle Fig.~\ref{missed-bug}. Lastly, the tests we used to find mixing bugs (\textsection\ref{missed}) are generated by \texttt{Memalloy}~\cite{Wickerson:2017:ACM:3009837.3009838} and \texttt{diy}~\cite{Alglave:2012:FWM:2205506.2205523}, although Fig.~\ref{mp_rc11} is not generated by today's tools~\cite{geeson2024weak}. In this case, we compare assembly program outcomes (\textsection\ref{challenge}) to find programs that induce bugs.

\section{Industry Impact}\label{imp}

In this section we cover our experience applying \mixtesting{} in industry. We worked closely with Arm's compiler teams to develop an \textit{atomics application binary interface} (ABI) that specifies the mappings of source-level atomics into AArch64 assembly sequences. As far as we know this is the industry's first public specification of an atomics ABI with an accompanying tool (\tttool{}) that can find bugs in non-compliant compilers. We summarize the specification as it is today, and refer the reader to the published document for updates~\cite{atomicsABI}.

\subsection{An ABI specification of Armv8 atomics}\label{abi}

The ABI is defined by a list of atomics mappings (1, below), accessed through compiler profiles that generate atomic instructions. Each mapping is correct if it satisfies a declarative statement of atomics ABI compatibility (2) when mix tested on a large sample of the test space (3).

\subsubsection{Listing atomics mappings}\label{one}: We test atomics mappings produced by the compiler profiles in LLVM and GCC that use \texttt{-march=armv8+\{lse|rcpc|rcpc3|lse128\}/armv8.4-a}. Since architecture sub-versions such as \texttt{-march=armv8.1-a} imply some of these flags they are omitted.

\begin{example}
Table.~\ref{mapping} shows how a 32-bit integer exchange maps to either a compare-and-swap sequence when Armv8.0 is selected or a swap instruction (\texttt{SWP}) if the Armv8.1-a is selected.

\begin{table*}[h]
  \caption{An exchange maps to a compare-and-swap loop or a \texttt{SWPL} instruction.}
  \begin{tabular}{|l|l|l|}
  \hline
  \textbf{Atomic Operation} & \textbf{Compiler Profile} & \textbf{Assembly Sequence}\\
  \hline
  \texttt{atomic\_exchange(loc,val,release)} & Base (\texttt{-march=armv8}) & 
  \texttt{\phantom{lbl:} MOV W2, \#val}\\
  & & \texttt{lbl: LDXR  W4,[loc]}\\
  & & \texttt{\phantom{lbl:} STLXR W3, W2,[loc]}\\
  & & \texttt{\phantom{lbl:} CBNZ W3,lbl}\\\cline{2-3}
                                        & \texttt{+lse} &
  \texttt{MOV W2, \#val}\\
  & & \texttt{SWPL W2, W4,[loc]}\\
  \hline
  \end{tabular}
  \label{mapping}
\end{table*}
\end{example}

\subsubsection{Statement of ABI compatibility}\label{two}: \emph{A compiler that implements the stated mappings is ABI-Compatible with respect to other compilers that implement the ABI.}

In other words, given a set of compiler profiles, a splitting function (Def.~\ref{strat}), and C/C++ litmus test set $S$, the mappings are correct with respect to $S$ if \mixtesting{} finds no mixing bugs (Def.~\ref{mbug}). This definition comes with the constraint that this is not a correctness guarantee, but rather a statement backed up by bounded testing. Verifying the compilation of concurrent programs under relaxed models is undecidable~\cite{10.1145/1706299.1706303}. Instead we test programs with a fixed initial state, loop unroll factor, and no recursion. The ABI does not make any statement about the compatibility of compilers outside the test bounds specified, the provided mappings are not exhaustive, the document makes no statement about the compatibility of optimised programs, nor any statements concerning the performance of compiled programs under the provided mappings. Nevertheless, a rigorous statement of ABI compatibility backed by an effective regression testing method and tool, is a significant improvement.

\subsubsection{Mix testing \ref{one} using \ref{two}}: We generate a number of concurrency tests, checking ABI compatibility of compilers that implement the mappings in the document. At the time of writing the mixing bugs reported in GCC are fixed, but not in LLVM.

\subsection{Using \tttool{} to test ABI compatibility}

By following the steps in \textsection\ref{tool} we mix test LLVM and GCC given compiler profiles and tests as input. We generate tests that involve patterns of C/C++ atomic operations, memory order parameters, barriers, control-flow and straight line code up to 5 threads in size. These tests are not exhaustive but aim to test atomic operations introduced in C/C++11. The ABI specifies mappings for C11 atomic operations for 8, 16, 32, 64, and 128-bit width accesses for both signed and unsigned integer types. Each atomic operation maps to multiple assembly sequences. Table.~\ref{bigt} defines all the combinations of test, compiler, and architecture under test.

\begin{table*}[h]
\centering
\caption{We test combinations of $\textnormal{C/C++ constructs} \times \textnormal{Acccess width/sign} \times \textnormal{Order} \times \textnormal{Arch}$.}
\begin{tabular}{|l|c|}
\hline
\textbf{C/C++ constructs:} & \texttt{(atomic operations|non-atomic operations}\\
& \texttt{ |barriers|control-flow|straight-line code)+}\\\hline
\textbf{Access width/sign:} & \texttt{(u)int(8|16|32|64)\_t}\\\hline
\textbf{Memory Order:} & \texttt{(relaxed|acquire|release|acquire-release|seq-cst)+}\\\hline
\textbf{Target Architecture:} &\texttt{(armv8|armv8+lse|armv8+rcpc|armv8+rcpc3|}\\
& \texttt{armv8+lse128 | armv8.4-a)+}\\
\hline
\end{tabular}
\label{bigt}
\end{table*}

We generated thousands of C/C++ litmus tests using \texttt{diy}~\cite{herdtools} and applied \tttool{} to get millions of AArch64 assembly litmus tests. We used \texttt{mshowhashes} to remove redundant compiled tests (see \textsection\ref{challenge}) and \texttt{herd}~\cite{herdtools} to search for mixing bugs. We parallelised~\cite{tange_2023_7668338} mix testing (with load balancing to reduce swap usage) on a 224 core ThunderX2 using 100GB runtime footprint and found no mixing bugs besides those we document in \textsection\ref{missed}. We do not auto-generate tests for all mappings in the ABI, since the \texttt{diy}~\cite{herdtools} generator does not support all read-modify-write operations, such as \texttt{fetch\_add}. We manually constructed tests with unsupported operations and applied \tttool{} to show there are no more mixing bugs in these cases.

\subsection{Variation of atomic mappings in practice}

There are many implementations of a given atomic operation in practice. Considering only the Armv8-A AArch64 backends of LLVM and GCC, there are up to 5 different mappings for each primitive, but many primitives also have mappings for each size and signedness. In addition, individual mappings were changed in compiler patches, but the changes were not consistently applied. As a result, LLVM and GCC are not currently interoperable, but specifying the ABI is one step towards addressing this. Altogether, the ABI specification fills over seventeen A4 pages, even with as much duplication removed as possible. Beyond this, there are also mappings used by proprietary compilers such as MSVC, which we have not yet considered. We expect that other compiler implementations for RISC-V, Intel x86, and IBM PowerPC have the potential for ABI mixing bugs too.

\subsection{Special cases}\label{special}
We detail two special cases that compilers should handle. These bugs were found by prior work~\cite{GeesonSmithCGO24}.

\begin{description}

\item[\textbf{Read-modify-write should preserve read}:] Exchange can map to \texttt{SWPL} instructions (Table.\ref{mapping}). However according to the Arm Architecture Reference Manual~\cite{Seal:2000:AAR:517257} \textit{instructions where the destination register is \texttt{WZR} or \texttt{XZR}, are not regarded as doing a read for the purpose of a \texttt{DMB LD} barrier}. The bug in Fig.~\ref{mp_rc11} arises since the effects of executing a \texttt{SWPL} may be reordered past the acquire fence (Fig.~\ref{swpbug}), we propose that compilers do not rewrite the destination register to be the zero register (\texttt{WZR}) in this case. This also applies to mappings using \texttt{LD<OP>} or \texttt{CAS}.

\begin{figure}[H]
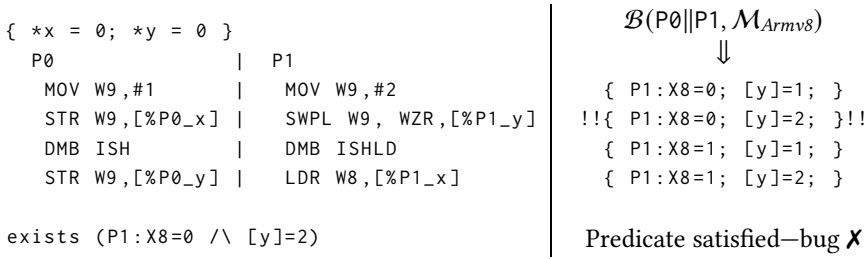

  \begin{tabular}{c | c}
\begin{lstlisting}[language=C, style=mystyle]
{ *x = 0; *y = 0 }
  P0              |  P1                    
   MOV W9,#1      |   MOV W9,#2            
   STR W9,[%P0_x] |   SWPL W9, WZR,[%P1_y]
   DMB ISH        |   DMB ISHLD           
   STR W9,[%P0_y] |   LDR W8,[%P1_x]
   
exists (P1:X8=0 /\ [y]=2)
\end{lstlisting} &
  \begin{tabular}{c}
  $\mathcal{B}(\texttt{P0} \Vert \texttt{P1},\mathcal{M}_{\mathit{Armv8}})$\\
  $\Downarrow$\\
\begin{lstlisting}[language=C, style=mystyle]
  { P1:X8=0; [y]=1; }
!!{ P1:X8=0; [y]=2; }!!
  { P1:X8=1; [y]=1; }
  { P1:X8=1; [y]=2; }
\end{lstlisting}\\\\
  Predicate satisfied---bug \xmark
  \end{tabular}
  \end{tabular}
  \caption{The compiled version of Fig.~\ref{mp_rc11}. The \texttt{SWPL} destination register is the \texttt{WZR} zero register.}
  \label{swpbug}
\end{figure}

\item[\textbf{Mutable const-qualified 128-bit data}:] Registers in AArch64 state hold 64-bit values. To load 128 bits atomically we must use a compare-and-swap loop (see Table.\ref{mapping128}) when Armv8.0 is selected. If \texttt{const}-qualified memory is marked read-only (and stored in, for example, \texttt{.rodata}) then executing the store-exclusive pair (\texttt{STXP}) instruction will crash the program. We propose that compliant implementations should mark \texttt{const}-qualified atomic locations as mutable. This also affects x86 code generation~\cite{constbugx86} of 64-bit access.

\begin{table*}[h]
  \caption{Some mappings for an 128-bit atomic load, in this case a compare-and-swap loop or an \texttt{LDP} instruction.}
  \begin{tabular}{|l|l|l|}
  \hline
  \textbf{Atomic Operation} & \textbf{Compiler Profile} & \textbf{Assembly Sequence}\\
  \hline
  \texttt{load(loc,relaxed)} & Base (\texttt{-march=armv8}) & 
  \texttt{lbl: LDXP X9, X10, [loc]}\\
  & & \texttt{\phantom{lbl:} STXP W3, X9, X10, [loc]}\\
  & & \texttt{\phantom{lbl:} CBNZ W3,lbl}\\\cline{2-3}
  & \texttt{+lse} &
  \texttt{LDP W2, W4, [loc]}\\
  \hline
  \end{tabular}
  \label{mapping128}
\end{table*}
\end{description}

\subsection{Sub-ABIs, ABI-islands, and the baseline ABI.}\label{island}

There is no one `true' ABI, but rather a specification that serves most purposes. The ABI we provide represents a \textit{baseline} specification for any implementation that aspires to be compatible across all versions of the Armv8 architecture. Ideally, mainstream implementations such as LLVM and GCC will adhere to this ABI in the future. This ABI does \textit{not} prevent implementors from creating their own ABI, whether it is a subset of the baseline (a \textit{sub-ABI}) or an altogether different set of mappings (a disjoint \textit{ABI-island}). A sub-ABI could induce mixing bugs on unsupported architectures (like in \textsection\ref{java}) and it would be up to the user of that sub-ABI to ensure such a situation cannot arise. Likewise, implementors may rely on an entirely different set of mappings that are disjoint from the baseline specification. Such an ABI-island would require similar restrictions to ensure correct execution. All ABI variants are of course relative to a baseline existing in the first place.

We observed that the absence of an explicit baseline led to the definition of implicit sub-ABIs. As architecture extensions (\textit{ie} fast new instructions) are introduced users quickly identify prospective mappings that offer performance improvements for their workloads. These sub-ABIs guide compiler development as they arise, but lacked ABI specification and testing until now. We provide a baseline ABI as guidance, firstly because mixing bugs have been introduced by accident (\textsection\ref{missed}). Secondly, there have been numerous attempts to optimise special atomic sequences (see \textsection\ref{special}), motivating the need to collect these cases together. Thirdly, engineers have been asked whether the same set of prospective mappings is correct by multiple different partners, and writing down the known cases helps rule out incorrect mappings. Lastly, the collective knowledge of atomics ABIs exists as a series of online discussions and web pages~\cite{mappings}, which are unfortunately outdated or have altogether disappeared (for instance when LLVM migrated from Phabricator to GitHub). We provide an ABI to help engineers and reduce the chances of mixing bugs arising in the future.

\subsection{Future ABI extensions}

The published atomics ABI~\cite{atomicsABI} contributes to an open ABI for the Arm Architecture. We hope that Arm's partners will submit requests to the ABI team\footnote{Please submit a request on GitHub~\cite{AABI} or contact \url{arm.eabi@arm.com}} so that more compilers can be validated and their mappings added to the ABI (if they differ from the current ABI). Further, the ABI team may consider new mappings if future architectures introduce new atomic instructions.

\section{Related work}\label{rel}
Neither compiler testing, memory models, nor interoperability is new. Previous work focuses on each topic in isolation whereas we combine them. We summarise the relevant work and show how \mixtesting{} improves upon them.

\begin{description}

\item[\textbf{Compiler testing for sequential code:}] Testing the compilation of sequential code has found hundreds of bugs in the past~\cite{Yang,10.1145/2666356.2594334}. There are broadly two approaches: \textit{differential testing} (see CSmith~\cite{Yang}) and \textit{metamorphic testing} (see Orion~\cite{10.1145/2666356.2594334}). In each case, techniques such as fuzzing and mutation are used to find bugs using many C/C++ features---a large test surface. Testing the compilation of concurrency samples a tiny test surface, namely the mappings from C/C++ atomics to assembly. Finding concurrency-related bugs is harder, since they require specific conditions to arise. As such prior concurrency testing work~\cite{10.1145/3460319.3469079,GeesonSmithCGO24,cmmtest,Chakraborty} found between two and six bugs each owing to the challenges of generating good tests and observing all behaviour. Mix testing finds four concurrency-related bugs but strictly generalises concurrency testing with respect to a single compiler profile, finding bugs that prior techniques cannot. We note that sequential testing techniques target the entire surface of the C/C++ languages, while our work is focused on the small, yet critical, surface of C/C++ atomics. Mix testing could be applied more broadly to find (not necessarily concurrency-related) ABI issues between compilers, which we defer to future work. We do not do any fuzzing or mutation.

\item[\textbf{Compiler testing for concurrent code:}] There has been a wealth of work on testing since the advent of the C/C++ model by Boehm and Adve~\cite{10.1145/1375581.1375591}, Batty~\cite{Batty:PhD} and others. \v{S}ev\v{c}\'{\i}k led the way on a theory of sound optimisations~\cite{Sevcik:2011:SOS:1993498.1993534} and verified compilation~\cite{10.1145/2487241.2487248}. The \texttt{cmmtest}~\cite{cmmtest} and \texttt{validc}~\cite{Chakraborty} tools are based on this theory and represent early efforts to find bugs. These tools compare the \textit{executions} of source and compiled tests. The \texttt{C4} tool~\cite{10.1145/3460319.3469079} simplifies testing by comparing \textit{outcomes} of executions; this is amenable to testing as compiler engineers know it. Unfortunately, \texttt{cmmtest} and \texttt{C4} rely on hardware executions, which may not exhibit all architecturally allowed outcomes, if the hardware exhibits them at all. Further \texttt{validc} does not test compilation down to the assembly level and may miss bugs in target dependent optimisations. Geeson and Smith 2024~\cite{GeesonSmithCGO24} address these issues by parametrising testing over source \textit{and} architecture models. \telechat{} is the simplest tool, comparing outcomes of source and compiled programs under their respective models.

\item[\textbf{Relaxed memory models:}] In the beginning, Lamport~\cite{Lamport:1979:MMC:1311099.1311750} coined the term \textit{sequential consistency} (SC). A \textit{relaxed} model is one that removes one or more constraints on the SC model. Fig.~\ref{fig:SBexample} uses SC accesses but is tested using relaxed models of C/C++~\cite{simonrc11} and Armv8. Proposals of new C/C++ models provide soundness proofs and \textit{compilation schemes} that suggest correct atomic mappings for compilers to implement (see for instance, Fig. 9 of Lahav et. al~\cite{Lahav:2017:RSC:3062341.3062352}). Unfortunately, such proofs are quickly outdated, since compilers implement \textit{multiple} mappings that can change. Mix testing exposes this fact and the subsequent bugs that follow. Whilst we do not contribute any work on the models themselves, \mixtesting{} implies there may exist mixing cases in soundness proofs that have not been considered. Geeson and Smith~\cite{GeesonSmithCGO24} conduct large-scale differential testing of various C/C++ models, but we leave mixed implementation soundness proofs to further work.

\item[\textbf{Language interoperability:}] Interoperability is a long-standing concern of engineers deploying portable code. The state-of-the-art testing techniques~\cite{cmmtest,Chakraborty,10.1145/3460319.3469079} assume the \textit{whole} program is compiled at once, using one set of atomics mappings. Unfortunately, this is an unrealistic view as production code bases are often compiled separately. Perconti and Ahmed~\cite{10.1007/978-3-642-54833-8_8} call this the \textit{closed-world} assumption and contribute correctness theorems for verification purposes. Mix testing is a testing analogue of this idea that we applied to production compilers and found mixing bugs. Our work is closer to \textit{combinatorial interaction testing} (CIT)~\cite{10.1145/4372.4375} that samples the test space to reduce the explosion of possible test parameters. CIT differs in that it constructs tests by exploiting the \textit{orthogonality} of test parameters whereas our choice of test input is coupled to each compiler profile under test. Since each atomic operation is implemented as a compiler intrinsic (whose code generation can change) that is accessed through compiler flags (that also change), we rely on expertise to pick these parameters. We document a number of practical (\textsection\ref{do}), theoretical (\textsection\ref{exponent}), and knowledge-based (\textsection\ref{abi}) techniques to reduce the complexity of \mixtesting{}.
\end{description}

\section{Conclusions and Further Work}\label{fut}

We present the \emph{\mixtesting{}} technique and the \tttool{} tool for testing the compilation of concurrent programs that mix atomic implementations. We explore the \mixtesting{} idea (\textsection\ref{practice}), technique and tool design, and problems we faced during implementation (\textsection\ref{impl}) with a reproducible artifact (see the Data-Availability statement below). We define a special kind of concurrency bug: the mixing bug (Def.~\ref{mbug}) and found four previously unknown mixing bugs in LLVM and GCC (\textsection\ref{missed}). We explore the exponential complexity of \mixtesting{} (\textsection\ref{exponent}) and practical ways to reduce the test generation (\textsection\ref{do}) and simulation penalty (\textsection\ref{challenge}).  We found a bug missed by the state-of-the-art on the same inputs (\textsection\ref{missed}). We expose the scale of testing the compilation of concurrency, as a problem that cannot be addressed by testing atomics mappings in isolation, but rather by testing in the presence of exponentially my many choices of mappings. This holds both now and in the future, as long as there are compatible atomics mappings in compilers. 

We show how the complexity of \mixtesting{} can be reduced in practice through our industry experience, notably by publishing an atomics application binary interface (ABI - \textsection\ref{abi}) for Armv8 AArch64 atomics implementations. As far as we know, this is the industry's first publicly documented~\cite{atomicsABI} specification of an atomics ABI with an accompanying tool that finds bugs in non-compliant compilers. The ABI reduces the complexity of \mixtesting{} to a smaller (but still exponential) number of implementations to test and provides validation for Arm's partners who build against it. Whilst developing the ABI, we assisted with queries from Arm's partners (\textsection\ref{missed}) regarding the correct mixing of atomics.

\section*{Data-Availability Statement}
The software that supports this paper is available on Zenodo~\cite{mixartifact}.

\begin{DIFnomarkup}
\begin{acks}
We thank Earl Barr, Richard Grisenthwaite, Al Grant, Kyrylo Tkachov, Tomas Matheson, Sam Ellis, Ties Stuij, Nick Gasson, Arm’s Compiler Teams and Arm Architecture \& Technology Group for their feedback and assistance. This work was supported by the Engineering and Physical Sciences Research Council [EP/V519625/1]. This work was also supported by EPSRC project [EP/R006865/1]. The views of the authors expressed in this paper are not endorsed by Arm or any other company mentioned.
\end{acks}
\end{DIFnomarkup}

\bibliographystyle{ACM-Reference-Format}
\bibliography{sample-base}

\end{document}